\documentclass[review,onefignum,onetabnum]{article}

\usepackage{latexsym}
\usepackage{amssymb,amsbsy,amsmath,amsfonts,amssymb,amscd,amsthm}
\usepackage{subfigure}
\usepackage{graphicx}
\usepackage{hyperref}
\usepackage{dsfont}
\usepackage{mathtools}
\usepackage{pdfsync}
\usepackage{epsfig,epstopdf}
\usepackage{caption}
\usepackage{xcolor}
\usepackage{hyperref}

\newtheorem{rmrk}[]{Remark}

\newcommand{\beq}{\begin{equation}}
\newcommand{\eeq}{\end{equation}}
\newcommand{\beqa}{\begin{eqnarray}}
\newcommand{\eeqa}{\end{eqnarray}}

\numberwithin{equation}{section}
\numberwithin{prpstn}{section}
\numberwithin{ass}{section}
\numberwithin{rmrk}{section}

\title{Pattern formation within phenotype-structured chemotactic populations}

\author{Tommaso Lorenzi\thanks{Department of Mathematical Sciences ``G. L. Lagrange'', Politecnico di Torino, 10129 Torino, Italy (tommaso.lorenzi@polito.it)}
\and
Kevin J. Painter\thanks{Inter-university Department of Regional and Urban Studies and Planning, Politecnico di Torino, 10129 Torino, Italy (kevin.painter@polito.it)}
}

\begin{document}
\maketitle

\begin{abstract}
Populations can become spatially organised through chemotaxis autoattraction, wherein population members release their own chemoattractant. Standard models of this process usually assume phenotypic homogeneity, but recent studies have shed illumination on the inherent heterogeneity within populations: in terms of chemotactic behaviour, trait heterogeneity can range from the sensitivity to attractant gradients to the rate at which attractants are produced. We propose a framework that accounts for this heterogeneity, extending the standard Keller-Segel model to a non-local formulation in which the population is continuously structured across some phenotype state space. Focussing on autoattraction, we allow both the chemotactic sensitivity and the rate of attractant secretion to vary across the population and suppose members can switch between different phenotype states. We extend classical Turing-type linear stability analyses to determine the impact of phenotypic structuring on pattern formation, showing that the rate of switching influences both the critical condition for self-organisation and subsequent pattern dynamics. Scenarios in which the chemotactic sensitivity and attractant secretion are positively or negatively correlated are used to highlight the significance of these results.
\end{abstract}

\section{Introduction}

Numerous populations self-organise into aggregated groups, including single-celled organisms, embryonic cells, and animals. The processes that drive such groupings have been the source of significant interest, for theoretical, experimental, and field scientists. Of the various mechanisms, chemotaxis (i.e. movement guided by chemical concentration gradients) has received much attention for its ubiquity across the natural world. Chemotaxis plays an important role in the self-organisation of bacteria and other single-celled organisms into clusters and mounds, structures embryonic cells during morphogenesis, mediates immune cell guidance to infections, and can also lead to the grouping of various animals, e.g. through responses to pheromones \cite{painter2019mathematical}.

\begin{figure}[t!] 
\includegraphics[width=\textwidth]{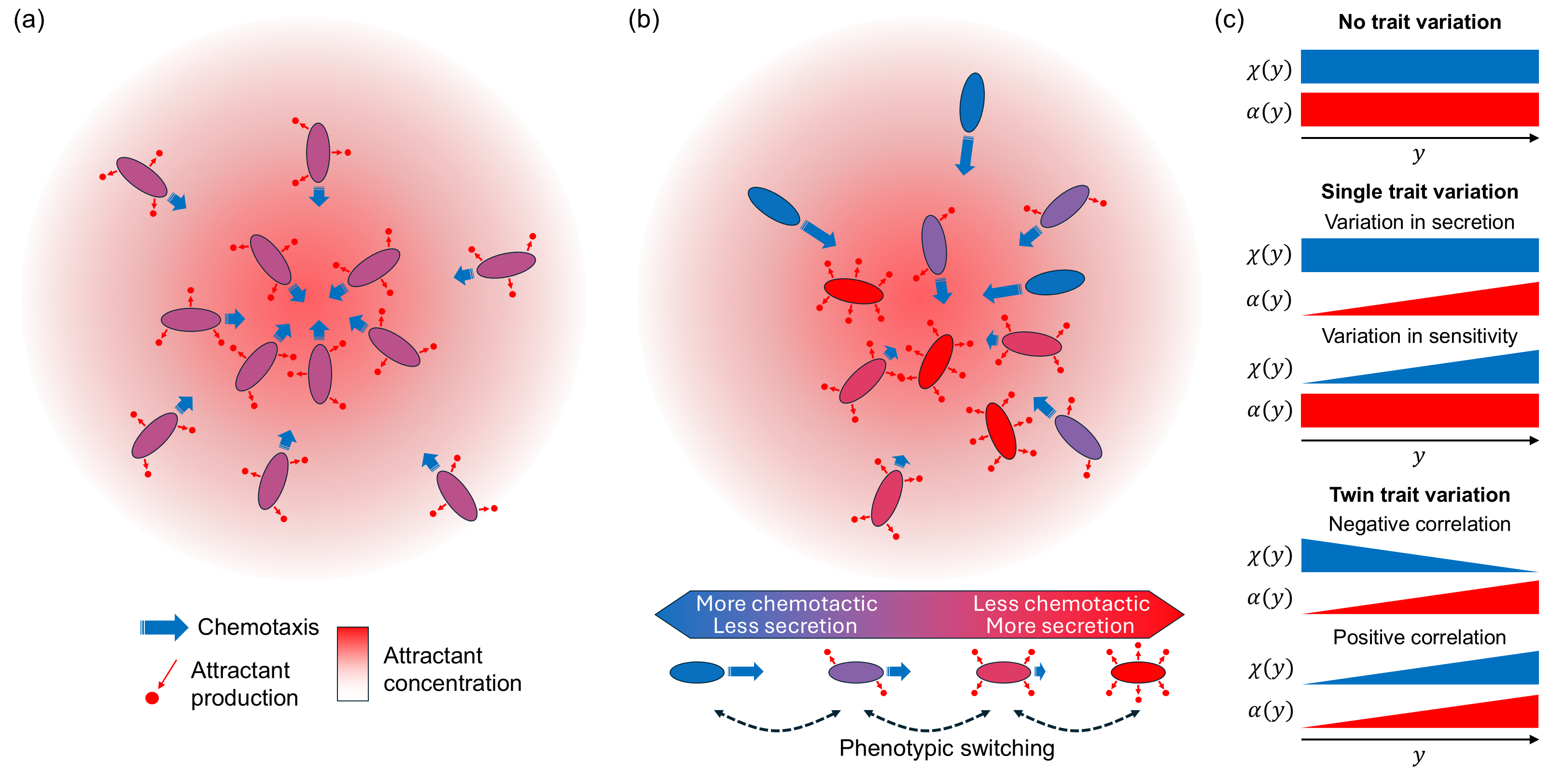}
	\caption{{\bf Self-organisation via positive feedback of autoattraction and related case study scenarios.} {\bf (a)} Autoattraction in a phenotypically homogeneous population, where each member secretes its own chemoattractant. {\bf (b)} Autoattraction in a phenotypically heterogeneous population, where members vary with respect to their rate of attractant secretion and their chemotactic sensitivity. In this schematic the traits are negatively correlated: highly chemotactic population members secrete little attractant and vice versa. Switching can occur between the phenotype states. {\bf (c)} Case study scenarios for variation in the chemotactic sensitivity, $\chi(y)$, and the rate of attractant secretion, $\alpha(y)$, considered here. From top to bottom: no trait variation, monotonic variation in a single trait (attractant secretion or chemotactic sensitivity), twin trait monotonic variation with the traits either negatively or positively correlated.}\label{figure1}
\end{figure}

Theoretical models for chemotaxis-driven self-organisation frequently rely on the framework of Keller and Segel \cite{keller1970initiation}, proposed more than 50 years ago. Stripped to an essential form, it comprises a coupled system of advection-diffusion-reaction equations for the density of a population, $\rho(t,x)$, and the concentration of a chemoattractant, $s(t,x)$, at position $x \in \Omega$ and time $t \in (0,\infty)$:
\beq
\label{eq:modelks}
\begin{cases}
\displaystyle{\partial_t \rho = {\rm div}\left(D_\rho \, \nabla_x \rho - \rho \, \chi \, \nabla_x s \right) + R(\rho) \, \rho,}
\\\\
\displaystyle{\partial_t s = D_s \Delta_x s + \alpha \rho - \eta \, s,}
\end{cases}
\quad (t,x) \in (0,\infty) \times \Omega.
\eeq
Self-organisation –- by which we mean the organisation of an (essentially) spatially uniform population into one or more aggregated groups/clusters –- has been studied intensively within this model \cite{painter2019mathematical}. Turing-type \cite{turing1952} linear stability analysis (see Appendix~\ref{appendix}) reveals pattern formation can be possible via the positive feedback of {\em autoattraction}: a population produces its own attractant (see Fig.~\ref{figure1}(a)). Particularly crucial, therefore, are the {\em chemotactic sensitivity}, $\chi$, and the {\em attractant secretion rate}, $\alpha$, and a threshold condition can be derived for the size of $\alpha \chi$ (see conditions~\eqref{eq:condinstclassic} and~\eqref{eq:condinstclassicwithgrowth} in Appendix~\ref{appendix}), a measure we define as the {\em autoattraction potential}. If the autoattaction potential is sufficiently high, the population density organises into one or more clusters and a range of spatiotemporal patterning phenomena can arise (see Figure \ref{figureA} in Appendix~\ref{appendix}).

An implicit assumption of~\eqref{eq:modelks} is phenotypic homogeneity: the population members are taken to have quasi-identical properties or {\it traits}, e.g. chemotactic sensitivity, attractant secretion rate, and so forth. An assumption of homogeneity is convenient for modelling, but must be viewed cautiously given the growing body of research that highlights within-population heterogeneity \cite{ackermann2015functional,carter2021epigenetic,keegstra2022ecological,keller2019unravelling,waite2018behavioral}. In terms of chemotactic movement, {\em E. coli} bacteria exposed to chemoattractant in a T-maze become sorted according to their chemotactic sensitivity, with the most chemotactic members penetrating deepest into the maze \cite{salek2019bacterial}. Variation in the secretion rate of an attracting substance can also occur. For instance, during their pheromone-mediated aggregation, tree-killing bark beetles considerably vary with respect to their level of pheromone secretion \cite{pureswaran2008high}. Summarising, across both cellular and animal populations, it is natural to suppose population variability across the components that mediate chemoattractant- or pheromone-induced aggregation. 

It is also natural to suppose that different traits could be positively or negatively correlated, that is, a population member with high chemotactic sensitivity may have a high or low attractant secretion rate. In cell populations, correlations could arise through interlinked signalling: a pathway that upregulates chemotaxis also upregulates chemoattractant release, as occurs in the response of {\em Dictyostelium discoideum} to the binding of cAMP to surface receptors \cite{weijer2004dictyostelium}. Negative correlations could also be a consequence of energetic constraints: as a behaviour incurs some energy cost, higher activity in one function may have to be balanced by lower activity in another \cite{keegstra2022ecological,ni2020growth}. 

Here we explore the impact of phenotypic variation on the capacity for self-organisation through chemotaxis. To retain simplicity and maintain generality, we consider an extension of the chemotaxis model~\eqref{eq:modelks} to a population that displays variability across chemotactic sensitivity and the rate of attractant secretion (see Section~\ref{sec:sec2}). We follow by extending classical pattern formation analyses to the extended model, for scenarios in which population growth kinetics are negligible (see Section~\ref{sec:pfnegpopgro}) or relevant (see Section~\ref{sec:pfpopgro}). We conclude with a summary of key findings and a discussion of future directions (see Section~\ref{sec:sec5}).

\section{A phenotype-structured chemotaxis model for self-organisation}
\label{sec:sec2}

Here we consider a possible generalisation of~\eqref{eq:modelks} to account for phenotypic heterogeneity, where the variability extends across two traits: chemotactic sensitivity and attractant secretion. As noted, these are the crucial determinants of whether a population can self-organise through autoattraction. 

\subsection{Model equations}

We assume a continuous structuring, with population members structured according to phenotype $y$, where $y$ is a continuous variable that spans some phenotype state space. For convenience, we set the phenotype state space to be one-dimensional and represented by a bounded interval $\mathcal{Y} \subset \mathbb{R}$. We assume that the dynamics of the local population density, $n(t,x,y)$, and the concentration of the chemoattractant, $s(t,x)$, are governed by the following coupled system of non-local advection-diffusion-reaction equations:
\beq
\label{eq:modelnspc}
\begin{cases}
\displaystyle{\partial_t n = {\rm div}\left(D_n \, \nabla_x n - n \, \chi(y) \, \nabla_x s \right) + \beta \, \partial^2_{yy} n + R(\rho) \, n, \quad y \in \mathcal{Y}},
\\\\
\displaystyle{\rho(t,x) = \int_{\mathcal{Y}} n(t,x,y) \, {\rm d}y,}
\\\\
\displaystyle{\partial_t s = D_s \Delta_x s + \int_{\mathcal{Y}} \alpha(y) \, n(t,x,y) \, {\rm d}y - \eta \, s,}
\end{cases}
\quad (t,x) \in (0,\infty) \times \Omega.
\eeq
Here the spatial domain $\Omega \subset \mathbb{R}^d$, with $d \geq 1$, is a bounded and connected set with smooth boundary $\partial \Omega$. We will refer to $n(t,x,y)$ as the \emph{phenotype density} and $\rho(t,x)$ as the \emph{(total) density}. Model~\eqref{eq:modelnspc} falls into the class of phenotype-structured partial differential equations (PS-PDEs) and we refer to \cite{lorenzi2024phenotype} for a review, including their formulation from an underlying random walk process (see also~\cite{lorenzi2025derivation}). For brevity, herein we refer to models~\eqref{eq:modelks} and~\eqref{eq:modelnspc} as the {\em unstructured chemotaxis model} and the {\em structured chemotaxis model}, respectively. 

In~\eqref{eq:modelnspc}$_1$\footnote{In this manuscript we use the notation $({\rm m})_i$ to refer to the equation on line $i$ for the group with number reference $({\rm m})$.}, the first term on the right-hand side takes into account undirected, random movement, which is (for simplicity) described through Fick’s first law of diffusion with diffusivity $D_n \in \mathbb{R}^+$, where $\mathbb{R}^+$ denotes the set of positive real numbers. The second term on the right-hand side corresponds to chemotactic movement (i.e. movement up the gradient of attractant) and the function $\chi(y)$, with $\chi: \mathcal{Y} \to \mathbb{R}^+_0$, where $\mathbb{R}^+_0$ denotes the set of non-negative real numbers, models the chemotactic sensitivity of individuals in the phenotype state $y$; note that by assuming this is non-negative we focus on scenarios where $s$ is an attractant for all phenotypes. The third term on the right-hand side of~\eqref{eq:modelnspc}$_1$ models phenotypic transitions across the population. We have modelled these here through a linear diffusion term with coefficient $\beta  \in \mathbb{R}^+_0$, which we will refer to as the {\em phenotype switching rate}. Naturally, one could consider other forms of phenotypic transitions -- such as directed towards a specific phenotype according to local densities or concentrations -- but as a first step we restrict to this relatively simple form. 

The final term on the right-hand side of~\eqref{eq:modelnspc}$_1$ is a non-local term that takes into account population growth kinetics (i.e. proliferation and/or death). The function $R(\rho)$, with $R : \mathbb{R}^+_0 \to \mathbb{R}$, is the local net growth rate of the population and we consider two regimes: negligible growth ($R\equiv0$) and non-negligible growth ($R\not\equiv0$). In the case of non-negligible growth, we let the function $R$ satisfy the following assumptions, so as to integrate the effect of density-dependent inhibition of growth (the cessation of growth at sufficiently high density):
\beq
\label{ass:Rnophdep}
R'(\cdot) < 0, \quad R(0) = \gamma, \quad R(\kappa)=0.
\eeq
Here $\gamma \in \mathbb{R}^+$ is the intrinsic growth rate of the population and $\kappa\in\mathbb{R}^+$ is the local carrying capacity for the (total) density.

In~\eqref{eq:modelnspc}$_3$, the first term on the right-hand side corresponds to Fickian diffusion, with diffusivity $D_s \in \mathbb{R}^+$ of the attractant. The second term models phenotype-dependent secretion of the attractant: attractant is secreted by individuals in the phenotype state $y$, at a rate described by the function $\alpha(y)$, with $\alpha: \mathcal{Y} \to \mathbb{R}^+_0$. Note that we restrict to non-negative functions for $\alpha(y)$, however one could consider scenarios in which certain phenotypes degrade attractant by allowing $\alpha(y)$ to range from positive to negative. The final term on the right-hand side of~\eqref{eq:modelnspc}$_3$ models natural decay of the attractant, which occurs at rate $\eta \in \mathbb{R}^+$. 

\subsection{Phenotype dependent functions}

We define the {\em trait-averaged attractant secretion rate}, $\left<\alpha\right>$, as
\beq\label{def:taasr}
\left<\alpha\right> = \frac{1}{|\mathcal{Y}|} \int_{\mathcal{Y}} \alpha(y) \, {\rm d}y\,,
\eeq
and the {\em trait-averaged chemotactic sensitivity}, $\left<\chi\right>$, as
\beq\label{def:tacs}
\left<\chi\right> = \frac{1}{|\mathcal{Y}|} \int_{\mathcal{Y}} \chi(y) \, {\rm d}y\,,
\eeq
where $|\mathcal{Y}|$ is the measure of set $\mathcal{Y}$. Recalling the criticality on chemotactic sensitivity and the attractant secretion rate for self-organisation, we will pay particular attention to the product of the above, $\left<\alpha\right>\left<\chi\right>$, as well as the {\em trait-averaged autoattractant potential}, $\left<\alpha \chi\right>$, defined as
\beq\label{def:taap}
\left<\alpha \chi\right> = \frac{1}{|\mathcal{Y}|} \int_{\mathcal{Y}} \alpha(y) \chi(y) \, {\rm d}y\,.
\eeq
Where possible, we retain general functional forms for $\alpha(y)$ and $\chi(y)$. However, to illustrate key results, we consider functions that are either constant or monotonic in the phenotype state $y$. Hence, for $\chi(y)$ we consider either  
\beq \label{eq:constantchi}
\chi(y) \equiv \chi_0
\eeq
or
\beq \label{eq:monotonechi}
\chi(y) = \chi_0 \, \hat{\chi}(y) \quad \mbox{ with } \quad \frac{1}{|\mathcal{Y}|}\int_{\mathcal{Y}}\hat{\chi}(y) \, {\rm d}y=1,
\eeq
where the function $\hat{\chi}(y)$ is monotonically increasing or monotonically decreasing on $\mathcal{Y}$. For both~\eqref{eq:constantchi} and~\eqref{eq:monotonechi} we have $\left< \chi \right> = \chi_0$ and the parameter $\chi_0 \in \mathbb{R}^+$ is therefore the trait-averaged chemotactic sensitivity. Likewise, for attractant secretion we consider either
\beq \label{eq:constantalpha}
\alpha(y) \equiv \alpha_0 
\eeq
or
\beq \label{eq:monotonealpha}
\alpha(y) = \alpha_0 \, \hat{\alpha}(y)  \quad \mbox{ with } \quad \frac{1}{|\mathcal{Y}|}\int_{\mathcal{Y}}\hat{\alpha}(y) \, {\rm d}y=1,
\eeq
where the function $\hat{\alpha}(y)$ is monotonically increasing or monotonically decreasing on $\mathcal{Y}$. For both~\eqref{eq:constantalpha} and~\eqref{eq:monotonealpha} we have $\left< \alpha \right> = \alpha_0$ and $\alpha_0 \in \mathbb{R}^+$ is thus the trait-averaged attractant secretion rate. 

These choices lead to five case study scenarios, as summarised by the schematics in Fig.~\ref{figure1}(c): \emph{no trait variation} (definitions~\eqref{eq:constantchi} and~\eqref{eq:constantalpha}); \emph{single trait variation}, either chemotactic sensitivity or attractant secretion (definitions~\eqref{eq:monotonechi} and~\eqref{eq:constantalpha} or definitions~\eqref{eq:constantchi} and~\eqref{eq:monotonealpha}); and \emph{twin trait variation} with the traits either positively or negatively correlated (definitions~\eqref{eq:monotonechi} and~\eqref{eq:monotonealpha} with functions $\hat{\chi}(y)$ and $\hat{\alpha}(y)$ that are monotone either in the same or in the opposite sense on $\mathcal{Y}$). Under these choices, via the Chebyshev integral inequality~\cite[p.~40]{mitrinovic1970analytic}, we note the following relationships between $\left< \alpha \right> \left< \chi \right>$ and $\left< \alpha \chi \right>$ defined via~\eqref{def:taasr}, \eqref{def:tacs}, and~\eqref{def:taap}:
\begin{itemize}
\item for no trait or single trait variation, 
\beq \label{eq:chebyshevnone}
\left< \alpha \chi \right> = \left< \alpha \right> \left< \chi \right> = \alpha_0 \chi_0;
\eeq
\item for positively correlated traits, 
\beq \label{eq:chebyshevpositive}
\left< \alpha \chi \right> > \left< \alpha \right> \left< \chi \right> = \alpha_0 \chi_0;
\eeq
\item for negatively correlated traits, 
\beq \label{eq:chebyshevnegative}
\left< \alpha \chi \right> < \left< \alpha \right> \left< \chi \right> = \alpha_0 \chi_0.
\eeq
\end{itemize}

As a final remark on forms for the phenotype dependent functions, for simulations we will set (without loss of generality) $\mathcal{Y} = (0,1)$, the attractant secretion rate as
\beq \label{eq:alphasimulation}
\alpha(y) = \alpha_0(p_\alpha+1)y^{p_\alpha},
\eeq
and the chemotactic sensitivity as
\beq \label{eq:chisimulation}
\begin{cases}
\chi(y) = \chi_0(p_\chi+1)y^{p_\chi} \, \mbox{ (if the two traits are positively correlated),}\\\\
\chi(y) = \chi_0(p_\chi+1)(1-y)^{p_\chi} \, \mbox{ (if the two traits are negatively correlated). } 
\end{cases}
\eeq
In the above, $p_{i}$ for $i \in\left\{\alpha,\chi\right\}$ represent nonlinearity coefficients. Setting $p_i=0$ results in no trait variation, while setting $p_i=1$ yields a linear variation. As $p_i\rightarrow \infty$ we approach extreme scenarios in which phenotype states other than $y=0$ and/or $y=1$ will have negligible attractant secretion and/or chemotactic sensitivity.

\subsection{Initial data and boundary conditions}

Initial data are set such that the following conditions hold
\beq
\label{eq:modelnsICspc}
\begin{cases}
\displaystyle{n(0,x,y) = n^0(x,y) \geq 0, \quad  \int_{\Omega}  \int_{\mathcal{Y}} n^0(x,y) \, {\rm d}y \, {\rm d}x = N^0>0}
\\\\
\displaystyle{s(0,x) = s^0(x) \geq 0,}
\end{cases}
\eeq
where $N^0$ defines the size of the population at $t=0$. Subsequently, we  define the {\em initial mean (total) density}, $\rho^0_{\rm m}$, and the {\em initial mean phenotype density}, $n^0_{\rm m}$, as 
\beq
\label{eq:ICs}
\rho^0_{\rm m} = \frac{N^0}{|\Omega|} \quad \mbox{and} \quad n^0_{\rm m} = \frac{\rho^0_{\rm m}}{|\mathcal{Y}|},
\eeq
where $|\Omega|$ is the measure of the set $\Omega$.

We complement system~\eqref{eq:modelnspc} with zero-flux boundary conditions on $\partial \Omega$, i.e. the following homogeneous Neumann boundary conditions
\beq
\label{eq:modelnsBCspc}
\nabla_x n(t,x,y)  \cdot \nu = 0 \;\; \forall \, (t,x,y) \in (0,\infty) \times \partial \Omega \times \overline{\mathcal{Y}}, \quad \nabla_x s  \cdot \nu = 0 \;\; \forall \, (t,x) \in (0,\infty) \times \partial \Omega,
\eeq
where $\nu$ is the unit normal to $\partial \Omega$ that points outwards from $\Omega$. We also impose zero-flux boundary conditions at the endpoints of $\mathcal{Y}$ for~\eqref{eq:modelnspc}$_1$, i.e. 
\beq
\label{eq:modelnsBCpheno}
\partial_y n(t,x,y) = 0 \;\; \forall \, (t,x,y) \in (0,\infty) \times \overline{\Omega} \times \partial \mathcal{Y}.
\eeq

\section{Pattern formation under negligible population growth}
\label{sec:pfnegpopgro}
We first investigate the case in which population growth kinetics can be considered negligible, focussing on a one-dimensional spatial scenario wherein $\Omega = (0,L)$ with $L \in \mathbb{R}^+$.

\subsection{Pattern formation analysis}
\label{sec:lsaR0}
In the corresponding unstructured chemotaxis model (i.e. system~\eqref{eq:modelks} with $R \equiv 0$), Turing-type linear stability analysis reveals a threshold condition on the autoattraction potential, $\alpha\chi$ -- see condition~\eqref{eq:condinstclassic} in Appendix~\ref{appendix}. To understand how phenotype structuring impacts on this condition, we extend the analysis to the structured chemotaxis model~\eqref{eq:modelnspc} with $R \equiv 0$. We notice that the full problem is analytically intricate, but illumination can be provided through certain relevant simplifications. Specifically, we consider first two asymptotic regimes for phenotype switching: {\em negligible} ($\beta \to 0$) or {\em fast} ($\beta\rightarrow\infty$), with respect to the timescales of movement and attractant dynamics. We then consider generic phenotype switching (i.e. $0<\beta<\infty$), where a similar analysis is possible but only under the case in which there is no variation in chemotactic sensitivity (i.e. when $\chi(y)$ is defined via~\eqref{eq:constantchi}). As demonstrated in Appendix~\ref{appendix:analysisR=0}, we formally find that the emergence of spatial patterns occurs:
\begin{itemize}
\item for $\beta \to 0$ when
\begin{equation}
\label{eq:condinstR0beta0}
\left< \alpha \chi \right> \rho^0_{\rm m} >  \eta \, D_n + \dfrac{m^2 \pi^2}{L^2}  \, D_n \, D_s \quad \text{for some } \; m \in \mathbb{N};
\end{equation}
\item for $\beta \to \infty$ when
\begin{equation}
\label{eq:condinstR0betainfty}
\left< \alpha \right> \left< \chi \right> \rho^0_{\rm m} > \eta \, D_n + \dfrac{m^2 \pi^2}{L^2}  \, D_n \, D_s  \quad \text{for some } \; m \in \mathbb{N};
\end{equation}
\item for $0<\beta<\infty$ and $\chi(y)$ defined via~\eqref{eq:constantchi} when
\begin{equation}
	\label{eq:condinstR0chiconst}
	 \left<\alpha\right>  \chi_0 \, \rho^0_{\rm m} >  \eta \, D_n + \dfrac{m^2 \pi^2}{L^2}  \, D_n \, D_s  \quad \text{for some } \; m \in \mathbb{N}.
\end{equation}
\end{itemize}

\begin{rmrk}
\label{rem:condsnogrowth}
As a consistency check, consider no trait variation (i.e. define $\chi(y)$ and $\alpha(y)$ via~\eqref{eq:constantchi} and~\eqref{eq:constantalpha}) and set $\mathcal{Y}=(0,1)$. Conditions~\eqref{eq:condinstR0beta0}, \eqref{eq:condinstR0betainfty}, and \eqref{eq:condinstR0chiconst} for pattern formation all reduce to 
$$
\alpha_0 \, \chi_0 \, \rho^0_{\rm m} > \eta \, D_n + \dfrac{m^2 \pi^2}{L^2}  \, D_n \, D_s  \quad \text{for some } \; m \in \mathbb{N},
$$
which is identical to the condition~\eqref{eq:condinstclassic} obtained for the unstructured chemotaxis model~\eqref{eq:modelks} with $R \equiv 0$. Moreover, for very large spatial domains (i.e. when $L \to \infty$) and very large populations (i.e. when $N^0 \to \infty$) such that the mean (total) density (i.e. $\rho^0_{\rm m}$ defined via~\eqref{eq:ICs}) remains constant, the above condition is simply $\chi_0 \, \alpha_0 \, \rho^0_{\rm m}> \eta D_n$. This recovers the standard idea that pattern formation can arise when the autoattraction potential overcomes attractant decay and cell dispersal.
\end{rmrk}

\subsection{Dynamics of pattern formation}
Pattern formation analysis reveals a subtle but significant difference to the pattern formation criterion as phenotypic switching ranges from negligible (cf. condition~\eqref{eq:condinstR0beta0}) to fast (cf. condition~\eqref{eq:condinstR0betainfty}), according to the trait-averaged chemotactic sensitivity, attractant secretion rate, and autoattraction potential (cf. definitions~\eqref{def:taasr}, \eqref{def:tacs}, and~\eqref{def:taap}). To highlight this difference we consider the case study scenarios illustrated in Fig.~\ref{figure1}(c), and for each of them we numerically solve the system~\eqref{eq:modelnspc} with $R \equiv 0$ subject to boundary conditions~\eqref{eq:modelnsBCspc}-\eqref{eq:modelnsBCpheno} and initial data that form a small perturbation of the uniform steady state \eqref{eq:barnbarsR0} (provided in Appendix~\ref{appendix:analysisR=0}). 

\subsubsection{No trait variation}

\begin{figure}[t!]    \includegraphics[width=\textwidth]{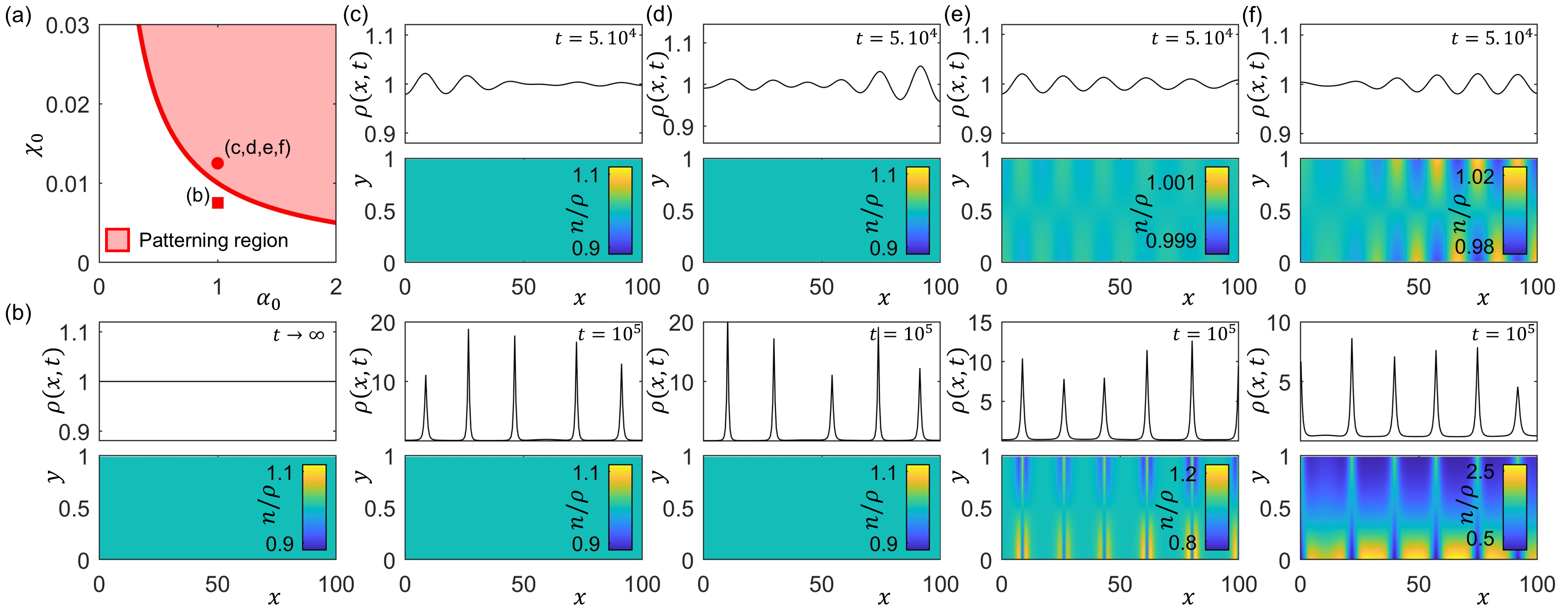}
		\caption{{\bf Pattern formation in the structured chemotaxis model~\eqref{eq:modelnspc} under negligible population growth with no trait/single trait variation.} {\bf (a)} Predicted patterning region in $(\alpha_0,\chi_0)$-space for the structured chemotaxis model~\eqref{eq:modelnspc} with $R\equiv 0$, under no trait variation (i.e. $\chi(y)$ and $\alpha(y)$ defined via~\eqref{eq:constantchi} and~\eqref{eq:constantalpha}), variation only in attractant secretion rate (i.e. $\chi(y)$ and $\alpha(y)$ defined via~\eqref{eq:constantchi} and~\eqref{eq:monotonealpha}), and variation only in chemotactic sensitivity (i.e. $\chi(y)$ and $\alpha(y)$ defined via~\eqref{eq:monotonechi} and~\eqref{eq:constantalpha}). {\bf (b)-(f)} Results from representative simulations of~\eqref{eq:modelnspc} with $R\equiv 0$. In each panel pair, the top panel displays the (total) cell density, $\rho(t,x)$, and the bottom panel displays the rescaled phenotype density, $n(t,x,y)/\rho(t,x)$, at specific times -- i.e. $t = 5 \cdot 10^4$ (top panel pair) and $t = 10^5$ (bottom panel pair). {\bf (b)} Outside the patterning region, numerical solutions evolve to distributions that are uniform in space and homogeneous in phenotype; here, $\alpha(y) \equiv 1$ and $\chi(y) \equiv 0.0075$, corresponding to the red square in (a). {\bf (c)-(f)} Simulations using parameter values that correspond to the red circle in (a): (c)~no trait variation with phenotype switching ($\alpha(y) \equiv 1$, $\chi(y) \equiv 0.0125$, and $\beta = 0.01$); (d) variation only in attractant secretion rate with phenotype switching ($\alpha(y) = 2y$, $\chi(y) \equiv 0.0125$, and $\beta = 0.01$); (e) variation only in chemotactic sensitivity with phenotype switching ($\alpha(y) \equiv 1$, $\chi(y) = 0.025y$, and $\beta = 0.01$); (f) variation only in chemotactic sensitivity with no phenotype switching ($\alpha(y) \equiv 1$, $\chi(y) = 0.025y$, and $\beta = 0$). For all numerical simulations we set $D_n=0.01$, $D_s=1$, $\eta=1$, $\Omega = (0,100)$, $\mathcal{Y}=(0,1)$, and initial data corresponding to a small perturbation of the uniform steady state \eqref{eq:barnbarsR0} (provided in Appendix~\ref{appendix:analysisR=0}), that is, $n_0(x,y) \equiv 1$, so that $\rho^0_{\rm m} = 1$, and $s_0(x)=1+r(x)$, where $r(x)$ is a small random perturbation of $\pm 10\%$.}\label{figure2}	
\end{figure}

Under no trait variation, the structured and unstructured chemotaxis models are (effectively) the same, hence offering a baseline for comparison. Defining $\chi(y)$ and $\alpha(y)$ via~\eqref{eq:constantchi} and~\eqref{eq:constantalpha}, the predicted patterning region across $(\alpha_0,\chi_0)$-space for ~\eqref{eq:modelnspc} with $R\equiv 0$ is displayed in Fig.~\ref{figure2}(a), highlighting the critical dependence on the trait-averaged chemotactic sensitivity and attractant secretion rates. When parameter values lie outside the predicted patterning region, numerical simulations reveal a population that becomes homogeneously distributed in phenotype and uniformly distributed in space (cf. Fig.~\ref{figure2}(b)). Conversely, if parameter values lie inside the predicted patterning region, we observe the emergence of spatially aggregated clusters. Unsurprisingly, given the lack of trait variation, we observe uniform spread across the phenotype states (cf. Fig.~\ref{figure2}(c)) and the (total) density $\rho(t,x)$ evolves qualitatively as for that of the unstructured chemotaxis model.

\subsubsection{Single trait variation}

We next consider variation in a single trait, either attractant secretion (hence, $\chi(y)$ and $\alpha(y)$ are defined via~\eqref{eq:constantchi} and~\eqref{eq:monotonealpha}) or chemotactic sensitivity (hence, $\chi(y)$ and $\alpha(y)$ are defined via~\eqref{eq:monotonechi} and~\eqref{eq:constantalpha}). For either case, the conditions under negligible \eqref{eq:condinstR0beta0} and fast \eqref{eq:condinstR0betainfty} phenotype switching are identical, suggesting that the rate of phenotype switching does not alter the fundamental capacity for pattern formation. Indeed, when the variation only applies to attractant secretion the same conditions hold also for generic phenotype switching (see condition~\eqref{eq:condinstR0chiconst}). 

Under definitions~\eqref{eq:constantchi} and~\eqref{eq:monotonealpha} or~\eqref{eq:monotonechi} and~\eqref{eq:constantalpha}, the trait-averaged autoattractant potential, $\left<\alpha \chi\right>$, defined via~\eqref{def:taap} coincides with the product of the trait-averaged attractant secretion rate, $\left<\alpha\right> = \alpha_0$, and the trait-averaged chemotactic sensitivity, $\left<\chi\right> = \chi_0$, defined via~\eqref{def:taasr} and~\eqref{def:tacs}. Consequently, these results logically extend the concept of a sufficient autoattractant potential obtained from the unstructured chemotaxis model~\eqref{eq:modelks}: provided that $\alpha_0 \chi_0$ is sufficiently large, self-organisation can occur. This dependence on averaged values implies that even population members with (at the level of their individual trait) a low or negligible autoattractant potential can still become spatially aggregated: Fig.~\ref{figure2}(d) shows the self-organisation of a population with varying attractant secretion rate, where we find low secretors to be equally represented within clusters.

While the pattern formation threshold remains the same, differences do emerge at the level of cluster distributions when the chemotactic sensitivity varies across the population. Members with a high sensitivity are found to dominate the centres of clusters, while those with lower sensitivity are found at the cluster periphery (cf. Fig.~\ref{figure2}(e)): the stronger chemotactic sensitivity of certain traits allows individuals expressing those traits to become more concentrated at the cluster centres. This structuring is amplified as the rate of phenotype switching becomes negligible (cf. Fig.~\ref{figure2}(f)). 

\subsubsection{Twin trait variation}

\begin{figure}
	\includegraphics[width=\textwidth]{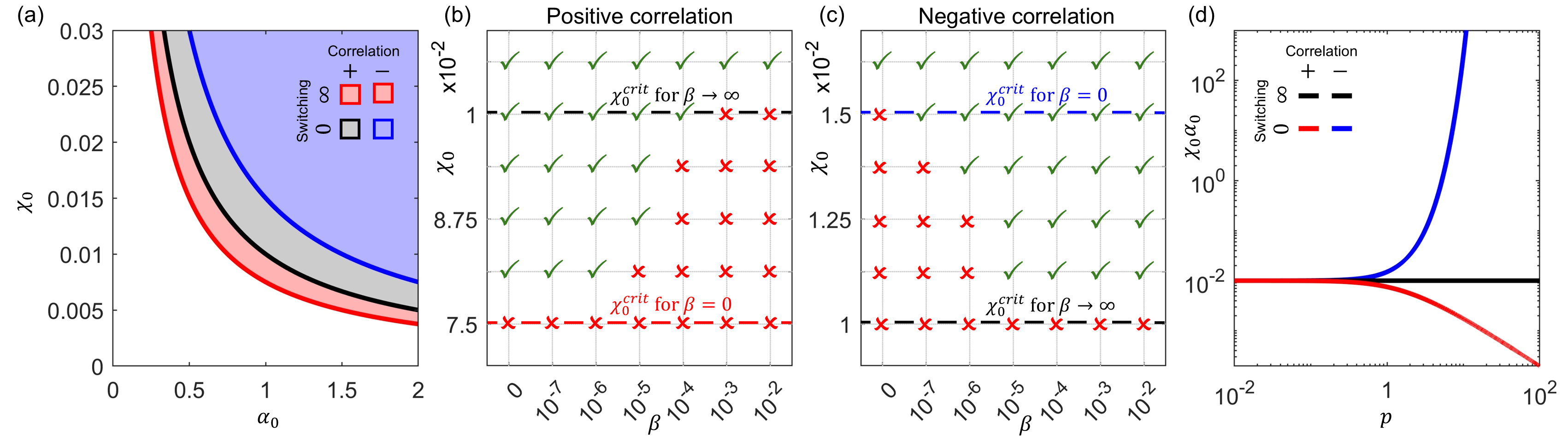}
	\caption{{\bf Pattern formation in the structured chemotaxis model~\eqref{eq:modelnspc} under negligible population growth with twin trait variation.} {\bf(a)} Predicted patterning region in $(\alpha_0,\chi_0)$-space under fast ($\infty$) or negligible ($0$) phenotype switching and positively ($+$) or negatively ($-$) correlated traits; $\alpha(y)$ and $\chi(y)$ are defined according to~\eqref{eq:alphasimulation} and ~\eqref{eq:chisimulation}, respectively, with $p_{\alpha}=p_{\chi}=1$. {\bf(b)-(c)} Summary of results of numerical simulations across $(\beta,\chi_0)$-space with $\alpha_0=1$ for positively correlated (b) and negatively correlated (c) traits, where ticks (crosses) mark points where pattern formation is (is not) numerically detected; $\alpha(y)$ and $\chi(y)$ are defined according to~\eqref{eq:alphasimulation} and ~\eqref{eq:chisimulation}, respectively, with $p_{\alpha}=p_{\chi}=1$. Dashed lines indicate the predicted critical chemotactic sensitivity from conditions~\eqref{eq:condinstR0beta0} and~\eqref{eq:condinstR0betainfty}. {\bf(d)} Changing threshold condition on $\alpha_0\chi_0$ for increasingly nonlinear trait variation; $\alpha(y)$ and $\chi(y)$ are defined according to~\eqref{eq:alphasimulation} and ~\eqref{eq:chisimulation}, respectively, with $p_{\alpha}=p_{\chi} = p$, under the four correlation-switching scenarios. For all numerical simulations we set $D_n=0.01$, $D_s=1$, $\eta=1$, $\Omega = (0,100)$, $\mathcal{Y}=(0,1)$, and initial data corresponding to a small perturbation of the uniform steady state \eqref{eq:barnbarsR0} (provided in Appendix~\ref{appendix:analysisR=0}), that is, $n_0(x,y) \equiv 1$, so that $\rho^0_{\rm m} = 1$, and $s_0(x)=1+r(x)$, where $r(x)$ is a small random perturbation of $\pm 10\%$.}
\label{figure3}
\end{figure}

We investigate variation across both attractant secretion and chemotactic sensitivity by defining $\chi(y)$ and $\alpha(y)$ via~\eqref{eq:monotonechi} and~\eqref{eq:monotonealpha}. In particular, we focus on the following four correlation-switching settings: $(+,0)$, $(+,\infty)$, $(-,0)$, and $(-,\infty)$, where $+/-$ indicates whether the correlation between the two traits is positive or negative (i.e. if the functions $\chi(y)$ and $\alpha(y)$ are monotone in the same sense or in the opposite sense on $\mathcal{Y}$), and $0/\infty$ indicates whether phenotype switching is negligible or fast. 

First, suppose a positive correlation where (without loss of generality) $\alpha(y)$ and $\chi(y)$ are both increasing on $\mathcal{Y}$. We have the relationship~\eqref{eq:chebyshevpositive} and, examining~\eqref{eq:condinstR0beta0} and~\eqref{eq:condinstR0betainfty}, we observe a distinct threshold on the size of $\alpha_0 \chi_0$ according to whether phenotype switching is negligible or fast. Specifically, a lower threshold is set when phenotype switching is negligible, suggesting that it is `easier' to achieve an aggregated state if switches in phenotype state are sufficiently rare. Next, consider a negative correlation where (without loss of generality) $\alpha(y)$ is increasing and $\chi(y)$ is decreasing on $\mathcal{Y}$. Here, we have the relationship~\eqref{eq:chebyshevnegative} and, consequently, the reverse scenario: a higher threshold is placed on $\alpha_0 \chi_0$ when phenotype switching is negligible. Significantly, these results suggest that the rate of phenotype switching plays an important role in determining whether pattern formation can occur, according to the relationship between the varying traits.

A visual demonstration of these results is provided in Fig.~\ref{figure3}(a), displaying the predicted patterning regions in ($\alpha_0,\chi_0)$-space for the various correlation-switching combinations considered here. Note that for these plots we have set $\mathcal{Y} = (0,1)$ and chosen linear phenotype dependent functions -- i.e. $\alpha(y)$ and $\chi(y)$ are defined according to~\eqref{eq:alphasimulation} and~\eqref{eq:chisimulation} with $p_{\alpha}=p_{\chi}=1$. Clearly, the patterning region expands under the $(+,0)$ setting, and shrinks under the $(-,0)$ setting. To explore the extent to which these results extend to other phenotype switching rates ($0 < \beta < \infty$), we compute numerical solutions of~\eqref{eq:modelnspc} with $R\equiv 0$ for different $(\beta,\chi_0)$-combinations, fixing $\alpha_0 =1$. Initial data are set as a small perturbation of the uniform steady state~\eqref{eq:barnbarsR0} (provided in Appendix~\ref{appendix:analysisR=0}). Under each form of correlation the analytical conditions provided by~\eqref{eq:condinstR0beta0} and~\eqref{eq:condinstR0betainfty} hold in the relevant regime for the phenotype switching rate $\beta$. Further, we observe a likely dependence on $\beta$ in general, such that the critical threshold monotonically varies between the values obtained for negligible and fast phenotype switching (cf. Figs.~\ref{figure3}(b),(c)). The precise nature of this relationship is left for future study. 
 
Finally, we explore the extent to which the threshold placed on $\alpha_0\chi_0$ changes under more extreme functional forms for the trait variation. Specifically, we set $\alpha(y)$ and $\chi(y)$ according to~\eqref{eq:alphasimulation} and ~\eqref{eq:chisimulation} with $p_{\alpha}=p_{\chi} = p$: as $p$ becomes large, we approach a `binary' scenario in which population members in phenotype states between 0 and 1 have negligible chemotactic sensitivity or attractant secretion rate. We observe that the different thresholds placed on $\alpha_0\chi_0$ according to ~\eqref{eq:condinstR0beta0} and~\eqref{eq:condinstR0betainfty} widen as the trait variation becomes more binary-like: increasingly relaxed for a positive correlation and increasingly severe for a negative correlation (cf. Fig.~\ref{figure3}(d)). 

\section{Pattern formation under population growth}
\label{sec:pfpopgro}
We now turn our attention to the full model, incorporating proliferation and death. As in Section~\ref{sec:pfnegpopgro}, we restrict to a one-dimensional spatial scenario, i.e. $\Omega = (0,L)$.

\subsection{Pattern formation analysis}
In the corresponding unstructured chemotaxis model (i.e. system~\eqref{eq:modelks} with $R \not\equiv 0$ that satisfies assumptions~\eqref{ass:Rnophdep}), Turing-type linear stability analysis once more reveals a threshold condition on the autoattraction potential, $\alpha\chi$ -- see condition~\eqref{eq:condinstclassicwithgrowth} in Appendix~\ref{appendix}. As in the above section, we now extend this analysis to the structured chemotaxis model~\eqref{eq:modelnspc} with $R \not\equiv 0$ under assumptions~\eqref{ass:Rnophdep}, considering the limiting cases in which $\beta \to 0$ and $\beta \to \infty$, and then the general case $0<\beta<\infty$ but when $\chi(y)$ is defined via~\eqref{eq:constantchi}. 

As observed in Appendix~\ref{appendix:analysisRneq0}, when $\beta \to 0$ algebraic complexity renders statement of a compact single minimal threshold (e.g. as condition~\eqref{eq:condinstR0beta0}) intractable. In fact, in this regime the eigenvalue problem that underlies the growth of linear instabilities driving pattern formation has a cubic nature, which raises the possibility of instabilities being either of stationary Turing type or of Turing-wave type (e.g. see \cite{krause2021modern}). In the latter case, emerging patterns would be expected to initially oscillate in time and space. Moreover, as previously, the terms appearing in pattern formation analysis depend on $\left<\alpha \chi \right>$, along with $\left<\alpha \right>$ and $\left<\chi \right>$. This leads us to again expect potentially distinct behaviours according to correlations between the rate of attractant secretion and chemotactic sensitivity, on which we expand further in Section~\ref{sec:csgrowth}.

On the other hand, compact conditions of the type of~\eqref{eq:condinstR0betainfty} and~\eqref{eq:condinstR0chiconst} are obtained in the other two cases (see Appendix~\ref{appendix:analysisRneq0}), where we find that the formation of spatial patterns occurs:
\begin{itemize}
\item for $\beta \to \infty$ when
\beq
\label{eq:condinstRcstbetainfty}
\left<\alpha\right> \left<\chi\right> \kappa  > \eta D_n + \frac{m^2 \pi^2}{L^2} D_n D_s + |R'(\kappa)| \, \left(D_s + \eta \frac{L^2}{m^2 \pi^2} \right) \, \kappa \quad \text{for some } \; m \in \mathbb{N};
\eeq
\item for $0<\beta<\infty$ and $\chi(y)$ defined via~\eqref{eq:constantchi} when
\begin{equation}
\label{eq:condinstRcstchiconst}
\left< \alpha \right> \, \chi_0 \, \kappa > \eta D_n + \frac{m^2 \pi^2}{L^2} D_n D_s + |R'(\kappa)| \, \left(D_s + \eta \frac{L^2}{m^2 \pi^2} \right) \, \kappa  \quad \text{for some } \; m \in \mathbb{N}.
\end{equation}
\end{itemize}

\begin{rmrk}
Note that, as in the observation made under negligible population growth (see Remark~\ref{rem:condsnogrowth}), if there is no trait variation (i.e. when $\chi(y)$ and $\alpha(y)$ are defined via~\eqref{eq:constantchi} and~\eqref{eq:constantalpha}) and $\mathcal{Y}=(0,1)$, then the conditions  for pattern formation derived in this section can all be found to reduce to 
$$
 \alpha_0 \, \chi_0 \, \kappa > \eta D_n + \frac{m^2 \pi^2}{L^2} D_n D_s + |R'(\kappa)| \, \left(D_s + \eta \frac{L^2}{m^2 \pi^2} \right) \, \kappa \quad \text{for some } \; m \in \mathbb{N}.
$$
This is identical to the condition~\eqref{eq:condinstclassicwithgrowth} obtained for the unstructured chemotaxis model~\eqref{eq:modelks} with $R \not\equiv 0$. 
\end{rmrk}

\subsection{Dynamics of pattern formation}
\label{sec:csgrowth}

\begin{figure}[p!]
	\includegraphics[width=\textwidth]{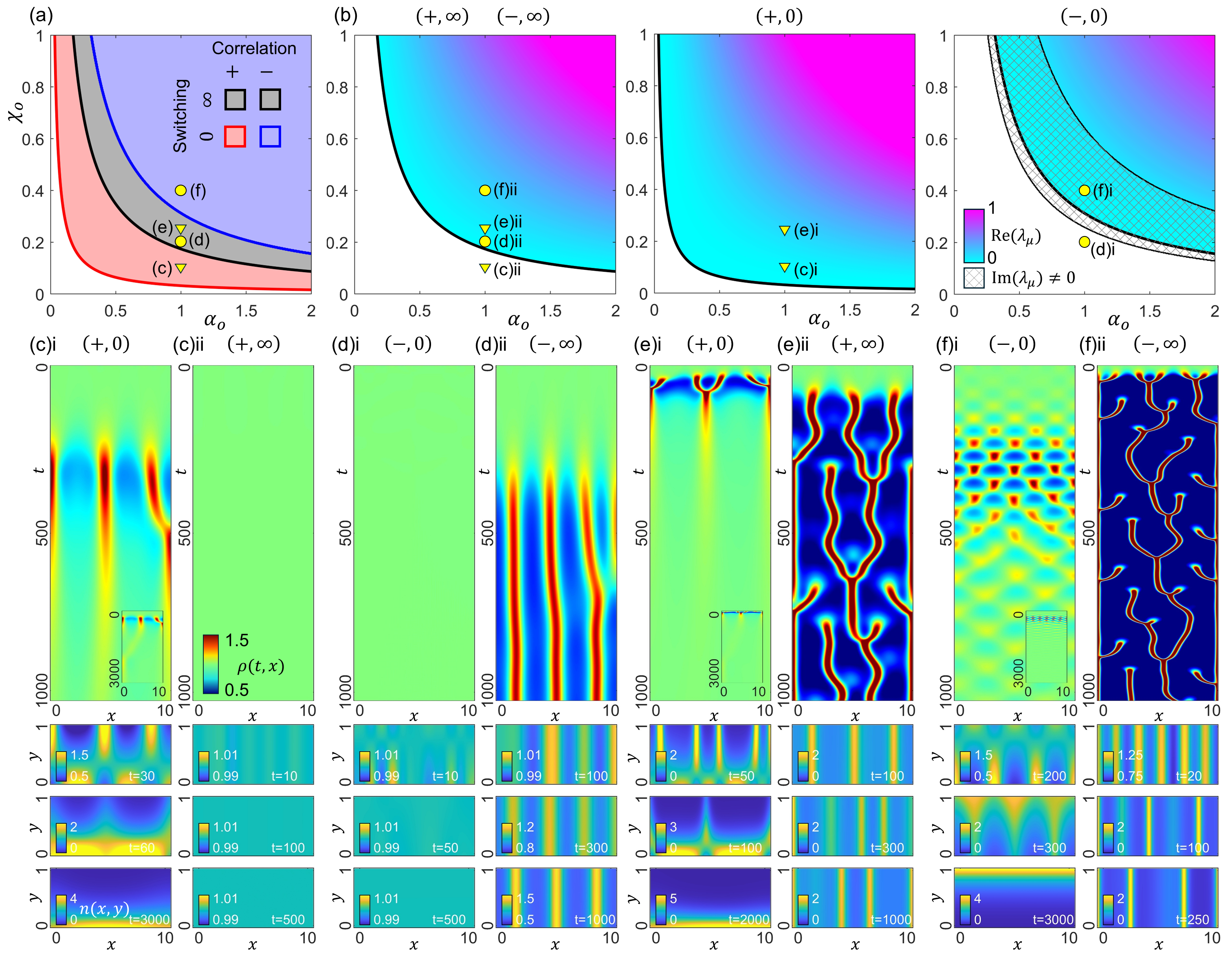}
	\caption{{\bf Pattern formation in the structured chemotaxis model~\eqref{eq:modelnspc} under population growth with negligible or fast phenotype switching.} {\bf (a)} Predicted patterning region in $(\alpha_0,\chi_0)$-space for the structured chemotaxis model~\eqref{eq:modelnspc}, with $R(\rho)$ defined via~\eqref{def:R}, under the four correlation-switching settings considered here. Note that indicated points mark parameter locations used to obtain the numerical solutions displayed in (c-f). {\bf (b)} Characteristics of the `most unstable' eigenvalue, $\lambda_\mu$, for the four correlation-switching settings in (a). Colour shade indicates the magnitude of its positive real part, and cross-hatching indicates whether there are imaginary parts. Imaginary parts only arise under the setting $(-,0)$, indicating a predicted bifurcation to patterns that oscillate in space and time. {\bf (c)-(f)} Representative numerical solutions of~\eqref{eq:modelnspc}, with $R(\rho)$ defined via~\eqref{def:R}, for (i) negligible ($\beta =0$) and (ii) `fast' ($\beta = 1$) phenotype switching; see text for details. Top panels display a kymograph of the (total) cell density $\rho(t,x)$ (colour scale as in (c)ii); bottom panels display the phenotype density $n(t,x,y)$ at the indicated times. In (c)-(f) we set: (c) $+$ve correlation with $(\alpha_0,\chi_0) = (1,0.1)$; (d) $-$ve correlation with $(\alpha_0,\chi_0) = (1,0.2)$; (e) $+$ve correlation with $(\alpha_0,\chi_0) = (1,0.25)$; (f) $-$ve correlation with $(\alpha_0,\chi_0) = (1,0.4)$. For all numerical simulations we set $D_n=0.01$, $D_s=1$, $\eta=1$, $\gamma=0.1$, $\kappa=1$, $\Omega = (0,10)$, $\mathcal{Y}=(0,1)$, and initial data corresponding to a small perturbation of the uniform steady state \eqref{eq:barnbarsRcst} (provided in Appendix~\ref{appendix:analysisRneq0}), that is, $n_0(x,y) \equiv 1$, so that $\rho^0_{\rm m} = 1$, and $s_0(x)=1+r(x)$, where $r(x)$ is a small random perturbation of $\pm 10\%$.}\label{figure4}	
\end{figure}

To test the implications of the pattern formation analysis, we explore pattern formation under twin trait variation, with the traits being positively or negatively correlated -- i.e. we define $\chi(y)$ and $\alpha(y)$ via~\eqref{eq:monotonechi} and~\eqref{eq:monotonealpha} and let the two functions be monotone in the same sense or in the opposite sense on $\mathcal{Y}$. For the calculations and simulations in this section, consistently with assumptions~\eqref{ass:Rnophdep}, we set 
\beq
\label{def:R}
R(\rho) = \gamma \, \left(1- \dfrac{\rho}{\kappa}\right).
\eeq
This definition can be viewed as an extension of a `logistic' form~\cite{verhulst1838notice} to a phenotype-structured model. Moreover we choose (without loss of generality) $\mathcal{Y}=(0,1)$, and use linear forms for the phenotype dependent functions -- i.e. $\alpha(y)$ and $\chi(y)$ are defined according to \eqref{eq:alphasimulation} and \eqref{eq:chisimulation} with $p_{\alpha}=p_{\chi}=1$. 

Predicted patterning regions in the $(\alpha_0,\chi_0)$-space for the structured chemotaxis model~\eqref{eq:modelnspc}, with $R(\rho)$ defined via~\eqref{def:R}, are displayed in Fig.~\ref{figure4}(a) for the correlation-switching combinations $(+,\infty)$, $(-,\infty)$, $(+,0)$, and $(-,0)$. The results reflect earlier findings in the absence of population growth: with negligible phenotype switching the patterning region is expanded under a positive correlation, but reduced for a negative correlation. To explore the resulting patterning we numerically solve the system~\eqref{eq:modelnspc} complemented with definition~\eqref{def:R}, under boundary conditions~\eqref{eq:modelnsBCspc}-\eqref{eq:modelnsBCpheno} and initial data that form a small perturbation of the uniform steady state~\eqref{eq:barnbarsRcst} (provided in Appendix~\ref{appendix:analysisRneq0}). At each parameter combination indicated in Fig.~\ref{figure4}(a), numerical solutions are computed with negligible ($\beta = 0$) and fast ($\beta = 1$) phenotype switching; note that $\beta>1$ yielded (essentially) identical behaviour, suggesting that $\beta=1$ can be regarded as suitably fast for this model setting. The predicted patterning regions are corroborated: for a positive correlation between attractant secretion and chemotactic sensitivity, an $(\alpha_0,\chi_0)$-combination that leads to aggregation when phenotype switching is negligible does not when phenotype switching is fast (cf. Fig.~\ref{figure4}(c)); conversely, for a negative correlation, an $(\alpha_0,\chi_0)$-pair that does not generate a pattern when phenotype switching is negligible can lead to aggregation when phenotype switching is fast (cf. Fig.~\ref{figure4}(d)).

Further, we observe significant differences with respect to the form of emerging patterns and long-term dynamics, according to the nature of the correlation between traits and the extent of phenotype switching. 

Firstly, we observe that emerging patterns oscillate in both space and time (cf. Fig.~\ref{figure4}(f)i) under the $(-,0)$ setting. As noted, the eigenvalue problem under negligible phenotype switching has a cubic nature (see equation~\eqref{eq:cubic} in Appendix~\ref{appendix:analysisRneq0}), which raises the possibility of a Turing-wave type instability. To investigate whether this is the case, we numerically compute eigenvalues across the $(\alpha_0,\chi_0)$-space and examine the characteristics of the `most unstable' eigenvalue, $\lambda_\mu$, which is defined as the eigenvalue with largest real part, across all wavenumbers $k$ (cf. Fig.~\ref{figure4}(b)). As $\alpha_0 \chi_0$ increases, we observe a bifurcation of stationary type for the $(-,\infty)$, $(+,\infty)$, and $(+,0)$ settings, but of Turing-wave type for the $(-,0)$ setting (cf. Fig.~\ref{figure4}(b)). Hence, linear stability analysis confirms the possibility of a Turing-wave type instability in the $(-,0)$ setting. 

Secondly, we notice that a prediction of pattern formation from the linear stability analysis does not imply that a pattern is sustained over longer timescales. Accordingly, under negligible phenotype switching, for both a positive (cf. Figs.~\ref{figure4}(c)i,(e)i) and negative (cf. Fig.~\ref{figure4}(f)i) correlation, the initial formation of aggregates is followed by aggregate decay. The emerging long-term distribution is uniform in space but heterogeneous in phenotype, concentrated around the phenotype states that correspond to low chemotactic sensitivity\footnote{Note that this is consistent with the fact that these numerical solutions are obtained by setting $\beta=0$. The positive uniform-in-space steady state~\eqref{eq:barnbarsRcst} (provided in Appendix~\ref{appendix:analysisRneq0}), which is also homogeneous in phenotype, is unique for $\beta > 0$; however, when $\beta = 0$ there can be positive uniform-in-space steady states that are not homogeneous in phenotype.}. This reveals a hidden `cost' to chemotactic aggregation, which can be attributed to the growth term: i) emergence of aggregates leads to clusters with peak densities that exceed the local carrying capacity; ii) these peaks are dominated by the most chemotactic population members, that subsequently experience higher rates of loss; iii) the steady elimination of the most chemotactic members diminishes the average chemotactic ability across the population, and eventually aggregates cannot be sustained. Phenotype switching, however, allows clusters to remain over time (cf. Figs.~\ref{figure4}(d)ii,(e)ii,(f)ii): while higher rates of loss still occur within clusters,  switching allows a continuous transition across traits to maintain a sufficient number of strongly chemotactic population members. We note that fast phenotype switching leads to a variety of dynamics that range from stable clusters (cf. Fig.~\ref{figure4}(d)ii) to aperiodic clusters that merge and emerge over time (cf. Figs.~\ref{figure4}(e)ii,(f)ii). These patterns bear a striking similarity to those observed in classic chemotaxis models~\cite{painter2011spatio} (see also Fig.~\ref{figureA} in Appendix~\ref{appendix}), reinforcing the notion that the dynamics of the structured chemotaxis model converge to those of the unstructured chemotaxis model under fast phenotype switching. 

\begin{figure}[p!]
	\includegraphics[width=\textwidth]{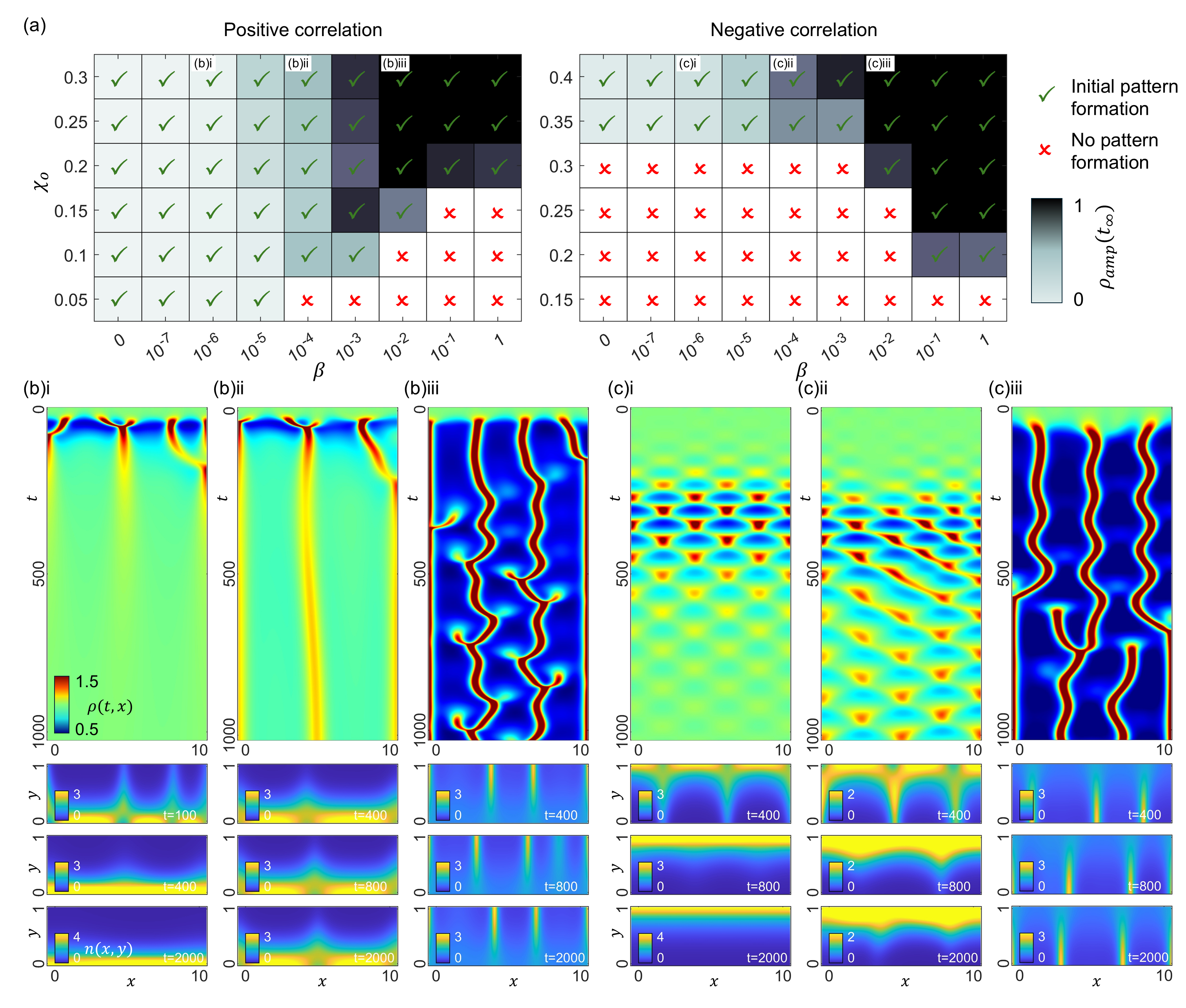}
	\caption{{\bf Pattern formation in the structured chemotaxis model~\eqref{eq:modelnspc} under population growth with generic phenotype switching.} {\bf (a)} Summary of results of numerical simulations across $(\beta,\chi_0)$-space for positively (left) and negatively (right) correlated traits, showing short-term and long-term pattern characteristics. Ticks (crosses) indicate whether initial pattern growth is (is not) detected. Shade indicates the amplitude of the pattern at the end of each simulation. {\bf (b)-(c)} Representative numerical solutions for a positive (b) and negative correlation (c) with $(\alpha_0,\chi_0)=(1,0.25)$ and $\beta=10^{-6}$ (i), $\beta=10^{-4}$ (ii), and $\beta=10^{-2}$ (iii). In (b)-(c) top panels display the kymograph for $\rho(t,x)$ (colour scale as in (b)i) and bottom panels display the phenotype density $n(t,x,y)$ at the times indicated. For all numerical simulations we set $D_n=0.01$, $D_s=1$, $\eta=1$, $\gamma=0.1$, $\kappa=1$, $\Omega = (0,10)$, $\mathcal{Y}=(0,1)$, and initial data corresponding to a small perturbation of the uniform steady state \eqref{eq:barnbarsRcst} (provided in Appendix~\ref{appendix:analysisRneq0}), that is, $n_0(x,y) \equiv 1$, so that $\rho^0_{\rm m} = 1$, and $s_0(x)=1+r(x)$, where $r(x)$ is a small random perturbation of $\pm 10\%$.}\label{figure5}	
\end{figure}

Finally, we examine the extent to which these observations hold for generic phenotype switching (i.e. when $0<\beta<\infty$). Specifically, we perform numerical simulations of~\eqref{eq:modelnspc} complemented with definition~\eqref{def:R} for different $(\beta,\chi_0)$-combinations, while fixing $\alpha_0 =1$. Each simulation is carried out until $t=10000$, or until the solution has evolved to a (numerically-determined) steady state. Fig.~\ref{figure5}(a) indicates short- and long-term pattern behaviour: short-term dynamics  refer to whether a pattern is detected to emerge from the initial perturbation of the uniform steady state; long-term refers to the eventual amplitude of the pattern, i.e. the quantity $\rho_{amp}(t_\infty) = \displaystyle{\max_{x\in \Omega}{\rho(t_\infty,x)}-\min_{x\in \Omega}{\rho(t_\infty,x)}}$, where $t_\infty$ represents the time at the end of the simulation. We observe a seemingly smooth transition between the behaviours observed under negligible and fast phenotype switching. Firstly, as $\beta$ increases we observe an increase or decrease in the critical chemotactic sensitivity that is required to trigger initial pattern formation, according to whether the correlation is positive or negative, respectively. Secondly, as $\beta$ increases we observe a growth in the amplitude of the long-term solution, from negligible when $\beta \approx 0$ to significant amplitudes when $\beta$ is relatively large. 

As specific examples, in Fig.~\ref{figure5}(b) we plot representative numerical solutions for increasing $\beta$ and a positive correlation, using the $(\alpha_0,\chi_0)$-combination of Fig.~\ref{figure4}(e). For small values of $\beta$ an initial pattern forms that subsequently decays, leading to a quasi-uniform distribution concentrated around phenotype states corresponding to low chemotactic sensitivity (cf. Fig.~\ref{figure5}(b)i), similar to that observed when $\beta = 0$ (cf. Fig.~\ref{figure4}(e)i). Sufficiently large values of $\beta$ lead to persisting large amplitude aggregates that undergo merging and emerging dynamics, in the manner of those observed in Fig.~\ref{figure4}(e)ii but with a noticeable variability along the phenotypic dimension (cf. Fig.~\ref{figure5}(b)iii). Middling $\beta$, on the other hand, gives rise to an intermediate pattern form: while the initial aggregates that emerge do decay and population members in phenotype states corresponding to a low chemotactic ability largely dominate, aggregates of a significant amplitude persist in time (cf. Fig.~\ref{figure5}(b)ii). 

In Fig.~\ref{figure5}(c) we plot representative solutions for increasing values of $\beta$ under a negative correlation, with the $(\alpha_0,\chi_0)$-combination of Fig.~\ref{figure4}(f). Again, we see behaviours that transition from those obtained when phenotype switching is negligible (cf. Fig.~\ref{figure4}(f)i) to fast (cf. Fig.~\ref{figure4}(f)ii). For intermediate values of $\beta$ here, patterns settle to a form that oscillates in space-time with a significant amplitude (cf. Fig.~\ref{figure5}(c)ii), while above some critical phenotype switching rate the initial oscillations in time disappear (cf. Fig.~\ref{figure5}(c)iii). 

\section{Discussion}
\label{sec:sec5}

In this work, we have extended the well-known Keller-Segel model for chemotaxis-driven self-organisation to account for continuous phenotypic structuring across certain key traits, namely attractant secretion and chemotactic sensitivity. These form the essential components of a positive feedback loop that allows a dispersed population to self-organise into aggregated groups \cite{keller1970initiation}: a process of autoattraction. By extending classical Turing linear stability analysis \cite{turing1952}, we have investigated how conditions for self-organisation change under phenotypic variation of a population. While Turing-type pattern formation analyses have been previously carried out for non-local PDE models of evolutionary dynamics~\cite{genieys2006adaptive,genieys2007dynamics} and PS-PDE models of chemically-interacting cells~\cite{ridgway2023motility}, as far as we are aware, the one presented here is the first such analysis for a chemotaxis model of this nature.

Under two general settings, negligible and non-negligible population growth, we observed a significant shift in the threshold for self-organisation, according to the rate at which phenotypic transitions occur. When population members can quickly switch between phenotypes, the well-known criticality placed on the chemotactic sensitivity and attractant secretion rate extends naturally: the product of trait-averaged chemotactic sensitivities and attractant secretion rates must be sufficiently large. When phenotype switching is negligible, however, it is the trait-averaged autoattractant potential that is the key determinant. While superficially similar, this has repercussions for traits that are either positively or negatively correlated: under a positive (resp. negative) correlation a lower (resp. higher) threshold is placed on the trait-averaged chemotactic sensitivity and attractant secretion rate.  

To place this in a practical context, consider population-level `effort' expended on certain functions. Chemotaxis (here measured by the model function $\chi(y)$) and chemoattractant synthesis/secretion (here measured by the model function $\alpha(y)$) each have a resource cost for an individual: taking {\em E. coli} as an example, chemotaxis places an energetic burden from the rotation of flagella to synthesis of chemoreceptors \cite{keegstra2022ecological}, as does the manufacture, transport, and secretion of chemoattractants. The trait-averaged sensitivity and secretion rate reflect the population-level costs of these activities, for a population initially dispersed across space and phenotype states. Within this context, if the traits are positively correlated, it can be viewed as more efficient to undergo minimal phenotype switching: a less stringent condition is found, meaning that aggregation can be achieved under lower trait-averaged values. Conversely, it is more efficient for members to switch frequently\footnote{We acknowledge that we have ignored any potential costs from switching for the purpose of this analogy.} when the traits are negatively correlated. 

At the individual-level effort, the model also provides insight into the extent to which populations can support `slackers', by which we mean those traits with a low autoattractant potential. Provided population-wide levels are sufficiently strong, self-organisation can occur: the individuals that save effort through contributing less than average still aggregate. However, if individuals with these traits become overrepresented, the overall autoattraction potential drops and self-organisation collapses at a population-wide level. We refer to \cite{pureswaran2008high} for some discuss on the potential energetic benefits of variable pheromone production rates, in the context of aggregation of tree-killing bark beetles.

Previous studies into the impact of phenotypic heterogeneity on chemotactic self-organisation have focussed on discrete phenotype models, in particular `binary phenotype' models in which the population is split between those exclusively performing chemotaxis and those that exclusively secrete attractant, e.g. \cite{macfarlane2022impact,tao2024nonlinear}. Under certain forms of phenotype switching, fast phenotype switching results in a convergence to the dynamics of the unstructured chemotaxis model \cite{macfarlane2022impact,painter2023phenotype}: effectively, the population converges such that it is governed by the average across the different traits. Here a fast phenotype switching appears to generate a similar phenomenon: the population becomes dispersed across all traits, the threshold criterion for pattern formation converges to that of the unstructured model, and spatiotemporal patterning dynamics are consistent. However, a clear divergence arises as switching events become rare. A fuller analytical investigation into the convergence between the structured and unstructured chemotaxis models would be of interest, for example within the context of global existence, a point of considerable attention for chemotaxis models \cite{bellomo2015toward}.

Under an aim of generality, we considered a non-specific variable that defines the phenotype. In practice, the phenotype will be linked to the internal state, such as intracellular signalling pathways for cell populations: we refer to \cite{mattingly2022collective,phan2024direct,ridgway2023motility} for related models where the phenotype-defining variable depends on some facet of intracellular signalling. We have also restricted to a diffusion-based model for phenotype switching, a reasonable simplification if phenotypic changes are driven by random fluctuations (e.g. stochastic changes in gene expression). Often, phenotype changes will be driven by environmental conditions \cite{schreiber2020environmental} and the model here could be easily adapted to include environmentally-driven phenotypic changes: for example, through the addition of advective terms that depend on environmental factors~\cite{almeida2024evolutionary,chisholm2016evolutionary,stace2020discrete}, such as the local attractant concentration~\cite{lorenzi2022trade}.

When growth was incorporated we obtained an even greater range of patterning, including complex merging/emerging dynamics and an instability that leads to emerging solutions that oscillate in space and time. The merging/emerging dynamics mirror those seen in classic unstructured chemotaxis models \cite{painter2011spatio} and can be observed in the regions where the structured model essentially behaves like the unstructured one (i.e. under fast phenotype switching). Oscillating solutions are found in the negative correlation case under negligible phenotype switching. Perhaps most notably, a prediction of spatial pattern formation does not necessarily imply long term patterning: over longer timescales, the amplitude of early forming patterns can decay to one with negligible spatial variation. The fact that predictions of pattern formation from linear stability analysis do not necessarily imply patterning at large time scales has been noted in other pattern formation models, e.g. \cite{krause2024turing}. Here it arises from a hidden cost to aggregation due to density-dependent inhibition of growth, penalising the aggregate via higher cell loss. 

Our current work has restricted phenotypic variation to the ability to perform chemotaxis and the secretion of the attractant: the focus on these two traits was natural, given their critical role for pattern formation. Naturally, other forms of trait variation could be considered, with proliferation being perhaps the most obvious: for {\em E. coli}, upregulated chemotaxis has been found to be associated with a downregulation in mitotic activity \cite{ni2020growth}; the go-or-grow hypothesis in cancer invasion postulates that cancer cells alternate between proliferative and migratory behaviour (for a recent review, see  \cite{thiessen2024go}). Including phenotype-dependent growth kinetics, therefore, would form a natural follow-on from the present work, e.g. giving certain traits greater proliferative ability or increased resistance against death. Intuitively, a higher growth potential within certain traits would confer a competitive advantage where, in the absence of other behaviours, the long term distribution will become concentrated to those traits (see \cite{lorenzi2024phenotype} for a review). The question of how that dominance is changed when self-organising dynamics are included would be of interest.

The study of self-organisation has formed a vast research topic for decades, dating to the groundbreaking model of \cite{turing1952} and considered in processes from morphogenesis \cite{landge2020pattern} to large-scale ecosystems \cite{rietkerk2021evasion}. Such systems are subject to heterogeneity: phenotypic variation across the cells in a developing tissue may alter individual rates of morphogen synthesis, degradation etc. Here we have explored how incorporating heterogeneity through continuous phenotype structuring impacts on patterning in a classic model, opening an avenue to study its role in other well-known models for self-organisation.

\section*{Acknowledgements} 
TL gratefully acknowledges support from the Italian Ministry of University and Research (MUR) through the grant PRIN 2020 project (No. 2020JLWP23) ``Integrated Mathematical Approaches to Socio-Epidemiological Dynamics'' (CUP: E15F21005420006) and the grant PRIN2022-PNRR project (No. P2022Z7ZAJ) ``A Unitary Mathematical Framework for Modelling Muscular Dystrophies'' (CUP: E53D23018070001) funded by the European Union{NextGenerationEU. TL and KJP are members of INdAM-GNFM.

\appendix
\renewcommand\thefigure{\thesection.\arabic{figure}}   
\section*{Appendix}

\section{Pattern formation analysis for the unstructured model~\eqref{eq:modelks}} 
\setcounter{figure}{0}  
\label{appendix}
For completeness, we recall the essentials of pattern formation analysis for the unstructured chemotaxis model~\eqref{eq:modelks} subject to: zero-flux boundary conditions on $\partial \Omega$, i.e. the following homogeneous Neumann boundary conditions
\beq
\label{eq:modelclassicBCs}
\nabla_x \rho(t,x)  \cdot \nu = 0 \;\; \forall \, (t,x) \in (0,\infty) \times \partial \Omega, \quad \nabla_x s  \cdot \nu = 0 \;\; \forall \, (t,x) \in (0,\infty) \times \partial \Omega
\eeq
where $\nu$ is the unit normal to $\partial \Omega$ that points outwards from $\Omega$; and initial data such that the following conditions hold
\beq
\label{eq:modelclassicICs}
\begin{cases}
\displaystyle{\rho(0,x) = \rho^0(x) \geq 0, \quad  \int_{\Omega} \rho^0(x) \, {\rm d}x = N^0 > 0}
\\\\
\displaystyle{s(0,x) = s^0(x) \geq 0.}
\end{cases}
\eeq

When $R\equiv 0$, positive uniform-in-space steady states of the system~\eqref{eq:modelks} are given by pairs $(\rho^0_{\rm m},\alpha \rho^0_{\rm m}/\eta)$ where $\rho^0_{\rm m}=N_0/|\Omega|$ represents the initial mean density of the population. We restrict to a one-dimensional spatial scenario, i.e. $\Omega = (0,L)$ with $L \in \mathbb{R}^+$, and consider small perturbations of such a steady state via the standard ansatz
\beq
\label{eq:ansatzclassic}
\rho(t,x) = \rho^0_{\rm m} + \tilde{\rho}  \exp\left(\lambda t\right) \, \varphi_k(x) \,, \qquad s(t,x) = \dfrac{\alpha \, \rho^0_{\rm m}}{\eta} + \tilde{s} \, \exp\left(\lambda t\right) \, \varphi_k(x)\,.
\eeq
Here, $\tilde{\rho}, \tilde{s} \in \mathbb{R}$ with $|\tilde{\rho}| \ll 1$ and $|\tilde{s}| \ll 1$, $\lambda \in \mathbb{C}$, $\{\varphi_k \}_{k \geq 1}$ are the eigenfunctions of the Laplace operator, acting on functions defined on $(0, L)$ and subject to homogeneous Neumann boundary conditions, indexed by the wavenumber $k = \dfrac{m \pi}{L}$ with $m \in \mathbb{N}$. Substituting the ansatz~\eqref{eq:ansatzclassic} into~\eqref{eq:modelks}, with $R\equiv0$, posed on $(0,L)$ and subject to the boundary conditions~\eqref{eq:modelclassicBCs}, retaining only linear terms one finds that the perturbations grow over time (that is, spatial pattern formation occurs) when 
\beq
\label{eq:condinstclassic}
\alpha \, \chi \, \rho^0_{\rm m} > \eta D_n + \frac{m^2 \pi^2}{L^2} D_n D_s  \quad \text{for some } \; m \in \mathbb{N}.
\eeq

On the other hand, in the case where $R\not\equiv 0$, under assumptions~\eqref{ass:Rnophdep} (i.e. encapsulating density-dependent inhibition of growth), the positive uniform-in-space steady state of the system~\eqref{eq:modelks} is given by the pair $(\kappa,\alpha \kappa/\eta)$. Applying linear stability analysis as when $R\equiv 0$, and using the fact that, under assumptions~\eqref{ass:Rnophdep}, the relation $-R'(\kappa)= \left|R'(\kappa)\right|$ holds, one finds that in this case spatial pattern formation occurs for 
\beq
\label{eq:condinstclassicwithgrowth}
\alpha \, \chi \, \kappa > \eta D_n + \frac{m^2 \pi^2}{L^2} D_n D_s + |R'(\kappa)| \, \left(D_s + \eta \frac{L^2}{m^2 \pi^2} \right) \kappa \quad \text{for some } \; m \in \mathbb{N}.
\eeq

A typical patterning region in the $(\alpha,\chi)$-space and representative examples of pattern formation dynamics are displayed in Fig.~\ref{figureA}, for the choice~\eqref{def:R} of $R(\rho)$. Below a critical threshold in the autoattraction potential, $\alpha \chi$, pattern formation is not possible, above the population organises into one or more aggregated clusters. Note that sufficiently high $\alpha \chi$ can lead to complex spatiotemporal phenomena, such as the merging and emerging of clusters (cf. Fig.~\ref{figureA}(b), right panel).

\begin{figure}[h!]
\begin{center}
\includegraphics[width=0.8\textwidth]{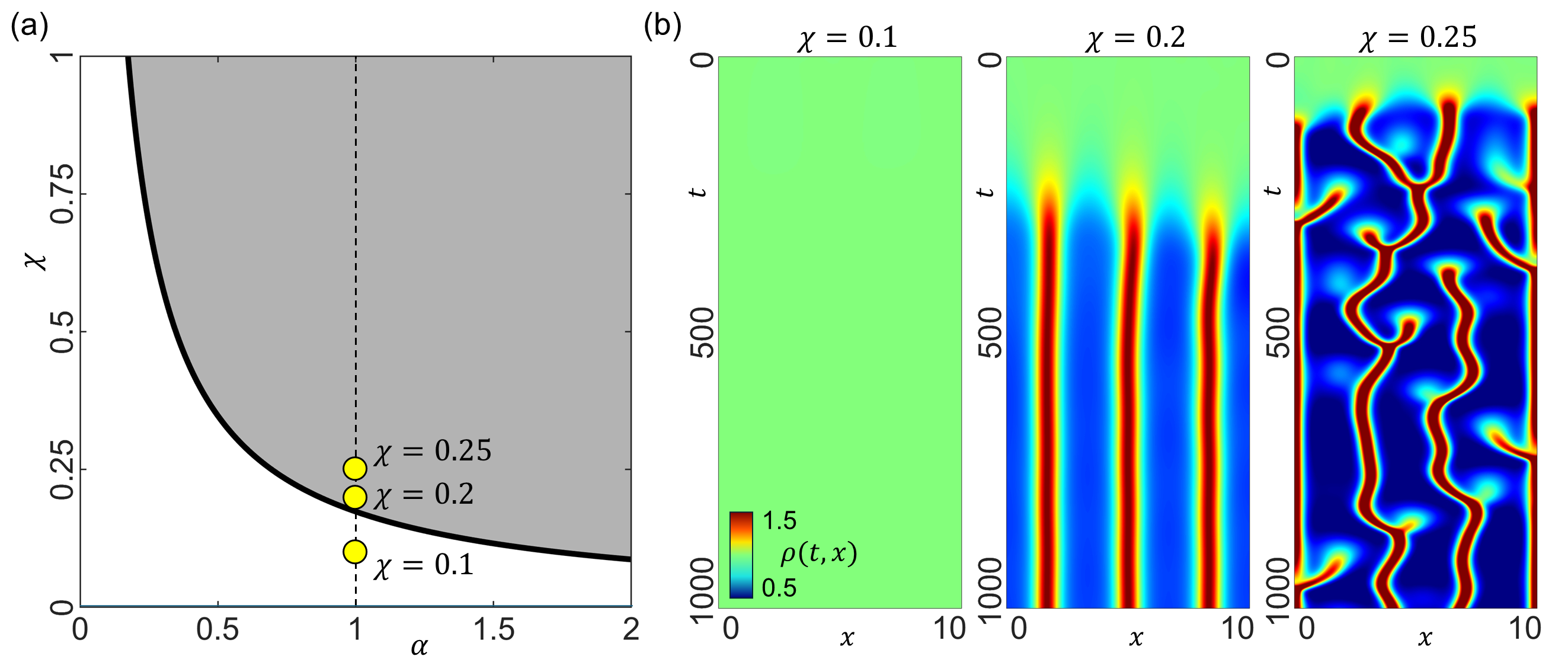}
\caption{{\bf Pattern formation in the unstructured chemotaxis model~\eqref{eq:modelks}.} {\bf (a)} Predicted patterning region in $(\alpha,\chi)$-space for the unstructured chemotaxis  model~\eqref{eq:modelks}, with $R(\rho)$ defined according to~\eqref{def:R}, showing a criticality in the magnitude of the autoattractant potential $\alpha\chi$. {\bf (b)} Kymographs of the cell density $\rho(t,x)$ (colour scale displayed in the left panel) when $\alpha=1$ and $\chi \in \{0.1, 0.2, 0.25\}$. For these we numerically solve~\eqref{eq:modelks}, complemented with definition~\eqref{def:R} and subject to the boundary conditions~\eqref{eq:modelclassicBCs} for $\Omega = (0,10)$. For all numerical simulations we fix $D_\rho=0.01$, $\gamma = 0.1$, $D_s=\alpha=\eta = \kappa = 1$, and set initial data that correspond to a small perturbation of the steady state $(\kappa,\alpha \kappa/\eta)$, that is, $\rho_0(x) \equiv 1$ and $s_0(x)=1+r(x)$, where $r(x)$ is a small random perturbation of $\pm 10\%$.}\label{figureA}
\end{center}
\end{figure}

\section{Pattern formation analysis for the structured model~\eqref{eq:modelnspc} with $R \equiv 0$}
\label{appendix:analysisR=0}
We carry out pattern formation analysis for the structured chemotaxis model ~\eqref{eq:modelnspc} with $R \equiv 0$, that is, the system
\beq
\label{eq:modelnspcR=0}
\begin{cases}
\displaystyle{\partial_t n = {\rm div}\left(D_n \, \nabla_x n - n \, \chi(y) \, \nabla_x s \right) + \beta \, \partial^2_{yy} n, \quad y \in \mathcal{Y}},
\\\\
\displaystyle{\partial_t s = D_s \Delta_x s + \int_{\mathcal{Y}} \alpha(y) \, n(t,x,y) \, {\rm d}y - \eta \, s,}
\end{cases}
\quad (t,x) \in (0,\infty) \times \Omega,
\eeq
which we complement with the boundary conditions~\eqref{eq:modelnsBCspc}-\eqref{eq:modelnsBCpheno}, and initial data such that conditions~\eqref{eq:modelnsICspc} hold. 

\paragraph{Positive uniform-in-space steady-state solutions}
Positive uniform-in-space steady-states of the system~\eqref{eq:modelnspcR=0} subject to conditions~\eqref{eq:modelnsBCspc}-\eqref{eq:modelnsBCpheno}, as well as initial data such that conditions~\eqref{eq:modelnsICspc} hold, are pairs $(\overline{n}(x,y), \overline{s}(x)) \equiv (\overline{n}(y), s^{\star})$, with $\overline{n} : \mathcal{Y} \to \mathbb{R}^+_0$ and $s^{\star} \in \mathbb{R}^+$ such that
\beq
\label{eq:modelnsststR0}
\begin{cases}
\displaystyle{\beta \, \partial^2_{yy} \overline{n}(y) = 0, \quad y \in \mathcal{Y}}
\\\\
\partial_y \overline{n}(y) =  0, \quad y \in \partial \mathcal{Y}
\\\\
\displaystyle{\int_{\mathcal{Y}} \overline{n}(y) \, {\rm d}y = \dfrac{N^0}{|\Omega|}}= \rho^0_{\rm m}
\\\\
\displaystyle{s^{\star} = \dfrac{1}{\eta} \, \int_{\mathcal{Y}} \alpha(y) \, \overline{n}(y) \, {\rm d}y}.
\end{cases}
\eeq

Solving the boundary-value problem~\eqref{eq:modelnsststR0}$_{1,2}$ for $\overline{n}(y)$ under the integral constraint~\eqref{eq:modelnsststR0}$_{3}$ and then substituting into~\eqref{eq:modelnsststR0}$_{4}$, recalling the definitions~\eqref{eq:ICs} for $\rho^0_{\rm m}$ and $n^0_{\rm m}$, as well as the definition~\eqref{def:taasr} for $\left<\alpha\right>$, we find
\beq
\label{eq:barnbarsR0}
\overline{n}(y) \equiv n^{\star} = \dfrac{1}{|\mathcal{Y}|} \, \dfrac{N^0}{|\Omega|} = \dfrac{\rho^0_{\rm m}}{|\mathcal{Y}|} =n^0_{\rm m}, \quad s^{\star} = \dfrac{1}{\eta} \dfrac{N^0}{|\Omega|} \dfrac{1}{|\mathcal{Y}|} \, \int_{\mathcal{Y}} \alpha(y) \, {\rm d}y = \frac{ \left<\alpha\right> \rho^0_{\rm m}}{\eta}.
\eeq
Hence, uniform-in-space steady-states are also homogeneous in phenotype.

\paragraph{Linear stability analysis of the positive uniform-in-space steady state}
Focussing on a one-dimensional spatial scenario wherein $\Omega = (0,L)$ with $L \in \mathbb{R}^+$, we consider small perturbations of the uniform steady state~\eqref{eq:barnbarsR0} via the ansatz
\beq
\label{eq:ansatzpertnsR0}
n(t,x,y) = n^{\star} + \tilde{\rho}  \exp\left(\lambda t\right) \, \varphi_k(x) \, \phi(y) , \qquad s(t,x) = s^{\star} + \tilde{s} \, \exp\left(\lambda t\right) \, \varphi_k(x),
\eeq
where $\tilde{\rho}, \tilde{s} \in \mathbb{R}$ with $|\tilde{\rho}| \ll 1$ and $|\tilde{s}| \ll 1$, $\lambda \in \mathbb{C}$, $\{\varphi_k \}_{k \geq 1}$ are the eigenfunctions of the Laplace operator, acting on functions defined on $(0, L)$ and subject to homogeneous Neumann boundary conditions, indexed by the wavenumber $k$, i.e.
\beq
\label{eq:wavenumbers}
k = \dfrac{m \pi}{L}, \quad m \in \mathbb{N},
\eeq
and the function $\phi : \mathcal{Y} \to \mathbb{R}$ satisfies the following normalisation condition
\beq
\label{eq:intphi}
\int_{\mathcal{Y}} \phi(y) \, {\rm d}y = 1.
\eeq
Substituting the ansatz~\eqref{eq:ansatzpertnsR0} into~\eqref{eq:modelnspcR=0} posed on $(0,L) \times \mathcal{Y}$ and subject to the boundary conditions~\eqref{eq:modelnsBCspc}-\eqref{eq:modelnsBCpheno}, retaining only linear terms yields
\beq
\label{newsyspert}
\begin{cases}
\displaystyle{\tilde{\rho} \, \beta \, \partial^2_{yy} \phi(y) - \tilde{\rho} \, \left(\lambda + D_n k^2 \right) \, \phi(y) = - \tilde{s} \, \chi(y) \, n^\star \, k^2, \quad y \in \mathcal{Y}}
\\\\
\partial_y \phi(y) = 0, \quad y \in \partial \mathcal{Y}
\\\\
\displaystyle{-\tilde{\rho} \, \int_{\mathcal{Y}} \alpha(y) \, \phi(y) \,  {\rm d}y + \tilde{s} \, \left(\lambda + \eta + D_s \, k^2 \right) = 0.}
\end{cases}
\eeq 
Note that integrating~\eqref{newsyspert}$_1$ over $\mathcal{Y}$ and imposing the boundary conditions~\eqref{newsyspert}$_2$ along with the normalisation condition~\eqref{eq:intphi} gives 
\beq
\label{newsyspertintcon}
\tilde{\rho} = \tilde{s} \, \dfrac{n^\star k^2}{(\lambda + D_n \, k^2)} \, \int_{\mathcal{Y}} \chi(y) \, {\rm d}y.
\eeq 
As explained in the body of the paper, we first consider the limiting scenarios in which $\beta \to 0$ and $\beta \to \infty$, and then turn to the general case $0<\beta<\infty$, where a similar analysis is possible but only when $\chi(y)$ is defined via~\eqref{eq:constantchi}.   

\subsection{Negligible phenotype switching}
\label{sec:annogrbeta0}
We first let $\beta \to 0$. Here, multiplying both sides of~\eqref{newsyspert}$_1$ by $\alpha(y)$, integrating over $\mathcal{Y}$ the resulting equation and imposing the boundary conditions~\eqref{newsyspert}$_2$, we formally find
$$
\tilde{\rho} \, \int_{\mathcal{Y}} \alpha(y) \, \phi(y) \,  {\rm d}y =  \tilde{s} \, \dfrac{n^\star k^2}{(\lambda + D_n \, k^2)} \int_{\mathcal{Y}} \, \alpha(y) \, \chi(y) \,  {\rm d}y.
$$
Substituting this into~\eqref{newsyspert}$_3$ and rearranging terms yields
$$
- \tilde{s} \, n^\star \, k^2 \, \int_{\mathcal{Y}} \alpha(y) \, \chi(y)  \, {\rm d}y \, + \, \tilde{s} \, \left(\lambda + \eta + D_s \, k^2 \right) (\lambda + D_n \, k^2) = 0
$$
from which we obtain the following quadratic equation for $\lambda \equiv \lambda(k^2)$
\beq
\label{eq:lambda}
\lambda^2 + b \, \lambda + c  = 0,
\eeq
where
$$
b \equiv b(k^2) = \left(D_n \, k^2 + D_s \, k^2 + \eta \right), \quad c \equiv c(k^2) = k^2 \, \left(D_n \, D_s \, k^2 + \eta \, D_n - n^\star \, \int_{\mathcal{Y}} \alpha(y) \, \chi(y)  \, {\rm d}y \right).
$$
For the considered perturbations to grow over time, and thus drive the steady state~\eqref{eq:barnbarsR0} unstable, it suffices that there exist some $k^2 \in \mathbb{R}^+$ such that $c(k^2)<0$. That is, recalling the expression~\eqref{eq:wavenumbers} of $k$, the expression~\eqref{eq:barnbarsR0} of $n^{\star}$, and the definition~\eqref{def:taap} of $\left< \alpha \chi \right>$, we formally find that the formation of spatial patterns occurs when condition~\eqref{eq:condinstR0beta0} is met, i.e. the condition
$$
\left< \alpha \chi \right> \rho^0_{\rm m} >  \eta \, D_n + \dfrac{m^2 \pi^2}{L^2}  \, D_n \, D_s \quad \text{for some } \; m \in \mathbb{N}.
$$

\subsection{Fast phenotype switching}
\label{sec:annogrbetainfty}

We next consider the opposite scenario (i.e. when $\beta \to \infty$). Here, \eqref{newsyspert}$_{1,2}$ along with the normalisation condition~\eqref{eq:intphi} formally give $\displaystyle{\phi(y) \equiv \dfrac{1}{|\mathcal{Y}|}}$. Inserting this along with the expression for $\tilde{\rho}$ given by~\eqref{newsyspertintcon} into~\eqref{newsyspert}$_3$ and rearranging terms yields
$$
- \tilde{s} \, n^\star \, k^2 \, \int_{\mathcal{Y}} \chi(y) \,  {\rm d}y \, \dfrac{1}{|\mathcal{Y}|} \, \int_{\mathcal{Y}} \alpha(y) \,  {\rm d}y \, + \, \tilde{s} \, \left(\lambda + \eta + D_s \, k^2 \right) (\lambda + D_n \, k^2) = 0,
$$
from which we obtain the quadratic equation~\eqref{eq:lambda} for $\lambda \equiv \lambda(k^2)$ with
$$
b \equiv b(k^2) = \left(D_n \, k^2 + D_s \, k^2 + \eta \right), \quad c \equiv c(k^2) = k^2 \, \left[\left(D_n \, D_s \, k^2 + \eta \, D_n\right) - n^\star \, \int_{\mathcal{Y}} \alpha(y) \, {\rm d}y \, \dfrac{1}{|\mathcal{Y}|} \int_{\mathcal{Y}} \chi(y) \, {\rm d}y \right].
$$
As previously, requiring that $c(k^2)<0$ and recalling the expression~\eqref{eq:wavenumbers} of $k$, the expression~\eqref{eq:barnbarsR0} of $n^{\star}$, and the definitions~\eqref{def:taasr} and~\eqref{def:tacs} of $\left< \alpha \right>$ and $\left< \chi \right>$, we formally find condition~\eqref{eq:condinstR0betainfty} for the formation of spatial patterns to occur, i.e. the condition
$$
\left< \alpha \right> \left< \chi \right> \rho^0_{\rm m} > \eta \, D_n + \dfrac{m^2 \pi^2}{L^2}  \, D_n \, D_s  \quad \text{for some } \; m \in \mathbb{N}.
$$

\subsection{Generic phenotype switching}
\label{sec:annogrchiconst}
A complete analysis for generic phenotype switching (i.e. $0 < \beta < \infty$) appears to be challenging, but the case where $\chi(y)$ is defined via~\eqref{eq:constantchi} is possible. Without loss of generality -- and to avoid cumbersome notation -- we set $\mathcal{Y}=(0,1)$, thereby $|\mathcal{Y}|=1$. Condition~\eqref{newsyspertintcon} reduces to
\beq
\label{newsyspertintconchiconst}
\tilde{\rho} = \tilde{s} \, \dfrac{n^\star k^2}{(\lambda + D_n \, k^2)} \, \chi_0.
\eeq
Moreover, solving the differential equation \eqref{newsyspert}$_1$ posed on $(0,1)$ under the condition~\eqref{newsyspertintconchiconst} and imposing the boundary conditions~\eqref{newsyspert}$_2$ at the endpoints $y=0$ and $y=1$, we find
\beq
\label{phichiconst}
\phi(y) \equiv 1.
\eeq
Substituting~\eqref{phichiconst} along with the expression for $\tilde{\rho}$ given by~\eqref{newsyspertintconchiconst} into~\eqref{newsyspert}$_3$ and rearranging terms yields
$$
- \tilde{s} \, n^\star \, k^2 \, \chi_0 \, \int_{0}^1 \alpha(y) \,  {\rm d}y \, + \, \tilde{s} \, \left(\lambda + \eta + D_s \, k^2 \right) (\lambda + D_n \, k^2) = 0,
$$
from which we obtain the quadratic equation~\eqref{eq:lambda} for $\lambda \equiv \lambda(k^2)$ with
$$
b \equiv b(k^2) = \left(D_n \, k^2 + D_s \, k^2 + \eta \right), \quad c \equiv c(k^2) = k^2 \, \left(D_n \, D_s \, k^2 + \eta \, D_n - n^\star \, \chi_0 \, \int_{0}^1 \alpha(y)  \, {\rm d}y \right).
$$
Once again it suffices that there exist some $k^2 \in \mathbb{R}^+$ such that $c(k^2)<0$. Recalling the expression~\eqref{eq:wavenumbers} of $k$, the expression~\eqref{eq:barnbarsR0} of $n^{\star}$, and the definition~\eqref{def:taasr} of $\left< \alpha \right>$, we find that the formation of spatial patterns occurs when condition~\eqref{eq:condinstR0chiconst} holds, i.e. when 
$$
	 \left<\alpha\right>  \chi_0 \, \rho^0_{\rm m} >  \eta \, D_n + \dfrac{m^2 \pi^2}{L^2}  \, D_n \, D_s  \quad \text{for some } \; m \in \mathbb{N}.
$$

\section{Pattern formation analysis for the structured model~\eqref{eq:modelnspc} with $R \not\equiv 0$}
\label{appendix:analysisRneq0}
We carry out pattern formation analysis for the structured chemotaxis model~\eqref{eq:modelnspc} with the function $R \not\equiv 0$ that satisfies assumptions~\eqref{ass:Rnophdep}.

\paragraph{Positive uniform-in-space steady-state solutions}
Positive uniform-in-space steady-states of the system~\eqref{eq:modelnspc} with $R \not\equiv 0$, subject to boundary conditions~\eqref{eq:modelnsBCspc}-\eqref{eq:modelnsBCpheno}, are pairs $(\overline{n}(x,y), \overline{s}(x)) \equiv (\overline{n}(y), s^{\star})$, with $\overline{n} : \mathcal{Y} \to \mathbb{R}^+_0$ and $s^{\star} \in \mathbb{R}^+$, such that
\beq
\label{eq:modelnsststRcst}
\begin{cases}
\displaystyle{\beta \, \partial^2_{yy} \overline{n}(y) + R(\rho^\star) \, \overline{n}(y) = 0, \quad y \in \mathcal{Y}}
\\\\
\partial_y \overline{n}(y) =  0, \quad y \in \partial \mathcal{Y}
\\\\
\displaystyle{\rho^{\star} =  \int_{\mathcal{Y}} \overline{n}(y) \, {\rm d}y}>0
\\\\
\displaystyle{s^{\star} = \dfrac{1}{\eta} \, \int_{\mathcal{Y}} \alpha(y) \, \overline{n}(y) \, {\rm d}y}.
\end{cases}
\eeq
Under assumptions~\eqref{ass:Rnophdep} on the net growth rate $R(\rho)$, solving the problem~\eqref{eq:modelnsststRcst}$_{1,2}$ for $\overline{n}(y)$ under the integral constraint~\eqref{eq:modelnsststRcst}$_{3}$ and then substituting into~\eqref{eq:modelnsststRcst}$_{4}$, recalling the definition~\eqref{def:taasr} for $\left<\alpha\right>$, we find
\beq
\label{eq:barnbarsRcst}
\overline{n}(y) \equiv n^{\star} = \dfrac{1}{|\mathcal{Y}|} \, \kappa, \quad s^{\star} = \dfrac{1}{\eta} \, \kappa\, \dfrac{1}{|\mathcal{Y}|} \, \int_{\mathcal{Y}} \alpha(y) \, {\rm d}y = \dfrac{\left< \alpha \right> \kappa}{\eta}.
\eeq
Hence, similarly to the case where population growth is negligible, the uniform-in-space steady-state is also homogeneous in phenotype. 

\paragraph{Linear stability analysis of the positive uniform-in-space steady state}
As in Appendix~\ref{appendix:analysisR=0}, we restrict to a one-dimensional spatial scenario, i.e. $\Omega = (0,L)$, and consider small perturbations of the steady state~\eqref{eq:barnbarsRcst} via the ansatz~\eqref{eq:ansatzpertnsR0}. Substituting this ansatz into system~\eqref{eq:modelnspc} posed on $(0,L) \times \mathcal{Y}$ and subject to the boundary conditions~\eqref{eq:modelnsBCspc}-\eqref{eq:modelnsBCpheno}, retaining only linear terms, using the normalisation condition~\eqref{eq:intphi} along with the fact that under assumptions~\eqref{ass:Rnophdep} the relation $-R'(\kappa)= \left|R'(\kappa)\right|$ holds, yields
\beq
\label{newsyspertRcst}
\begin{cases}
\displaystyle{\tilde{\rho} \, \beta \, \partial^2_{yy} \phi(y) - \tilde{\rho} \, \left(\lambda + D_n k^2 \right) \, \phi(y) = - \tilde{s} \, \chi(y) \, n^{\star} \, k^2 + \tilde{\rho} \, \left|R'(\kappa)\right| \, n^{\star}, \quad y \in \mathcal{Y}}
\\\\
\partial_y \phi(y) = 0, \quad y \in \partial \mathcal{Y}
\\\\
\displaystyle{-\tilde{\rho} \, \int_{\mathcal{Y}} \alpha(y) \, \phi(y) \,  {\rm d}y + \tilde{s} \, \left(\lambda + \eta + D_s \, k^2 \right) = 0.}
\end{cases}
\eeq 
Note that integrating equation~\eqref{newsyspertRcst}$_1$ over $\mathcal{Y}$ and imposing the boundary conditions~\eqref{newsyspertRcst}$_2$ along with the normalisation condition~\eqref{eq:intphi} gives
\beq
\label{newsyspertintconRcst}
\tilde \rho = \tilde{s} \, \dfrac{n^\star \, k^2}{(\lambda + D_n \, k^2 + \left|R'(\kappa)\right| \, \kappa)} \, \int_{\mathcal{Y}} \chi(y) \, {\rm d}y.
\eeq 
As in Appendix~\ref{appendix:analysisR=0}, we first consider the limiting scenarios in which $\beta \to 0$ and $\beta \to \infty$, and then turn to the general case $0<\beta<\infty$ but for $\chi(y)$ defined via~\eqref{eq:constantchi}. 

\subsection{Negligible phenotype switching}\label{sec:421}
When $\beta \to 0$, multiplying both sides of~\eqref{newsyspertRcst}$_1$ by $\alpha(y)$ and integrating over $\mathcal{Y}$ the resulting equation we formally obtain
$$
\tilde{\rho} \, \int_{\mathcal{Y}} \alpha(y) \, \phi(y) \,  {\rm d}y =  \tilde{s} \, \dfrac{n^\star k^2}{(\lambda + D_n \, k^2)} \int_{\mathcal{Y}} \, \alpha(y) \, \chi(y) \,  {\rm d}y \, - \, \tilde{\rho} \,  \dfrac{\left|R'(\kappa)\right| \, n^\star}{(\lambda + D_n \, k^2)} \int_{\mathcal{Y}} \, \alpha(y) \,  {\rm d}y.
$$
Moreover, inserting the expression for $\tilde \rho$ given by~\eqref{newsyspertintconRcst} into the right-hand side of the above equation yields 
$$
\tilde{\rho} \, \int_{\mathcal{Y}} \alpha(y) \, \phi(y) \,  {\rm d}y = \tilde{s} \, \dfrac{n^\star k^2}{(\lambda + D_n \, k^2)} \, \left(\displaystyle{\int_{\mathcal{Y}} \, \alpha(y) \, \chi(y) \,  {\rm d}y} - \dfrac{\left|R'(\kappa)\right| \, n^\star}{(\lambda + D_n \, k^2 + |\mathcal{Y}| \, \left|R'(\kappa)\right| \, n^\star)} \,  \int_{\mathcal{Y}} \chi(y) \, {\rm d}y \,  \int_{\mathcal{Y}} \, \alpha(y) \,  {\rm d}y\right).
$$
Substituting this into~\eqref{newsyspertRcst}$_3$ and rearranging terms gives
{\small
$$
- \tilde{s} \, n^\star \, k^2 \, \left(\displaystyle{\int_{\mathcal{Y}} \, \alpha(y) \, \chi(y) \,  {\rm d}y} - \dfrac{\left|R'(\kappa)\right| \, n^\star}{(\lambda + D_n \, k^2 + |\mathcal{Y}| \,\left|R'(\kappa)\right| \, n^\star)} \,  \int_{\mathcal{Y}} \chi(y) \, {\rm d}y \,  \int_{\mathcal{Y}} \, \alpha(y) \,  {\rm d}y\right) \, + \, \tilde{s} \, \left(\lambda + \eta + D_s \, k^2 \right) (\lambda + D_n \, k^2) = 0,
$$
}
from which, recalling the expression~\eqref{eq:barnbarsRcst} of $n^\star$ as well as the definitions~\eqref{def:taasr}, \eqref{def:tacs}, and~\eqref{def:taap} of $\left< \alpha \right>$, $\left< \chi \right>$, and $\left< \alpha \chi \right>$, we formally obtain the following cubic equation for $\lambda \equiv \lambda(k^2)$
\beq
\label{eq:cubic}
\lambda^3 + b \, \lambda^2 + c \, \lambda + d  = 0,
\eeq
where
\begin{eqnarray*}
&&b \equiv b(k^2) = 2 \, D_n \, k^2 + \eta + D_s \, k^2 + \left|R'(\kappa)\right| \, \kappa,
\\\\
&&c \equiv c(k^2) = \left(D_n \, k^2 + \eta + D_s \, k^2\right) \,  \left(D_n \, k^2 +  \left|R'(\kappa)\right| \, \kappa\right) +  \left(\eta + D_s \, k^2\right) \, D_n \, k^2 - k^2 \, \kappa \, \left<\alpha \chi \right>,
\\\\
&&d \equiv d(k^2) = D_n \, k^4 \left[ \left(\eta + D_s \, k^2\right) \,  \left(D_n + \dfrac{\left|R'(\kappa)\right| \, \kappa}{k^2}\right) - \,  \kappa \, \left<\alpha \chi \right> \right] \, + 
\\
&& \phantom{d \equiv d(k^2) =} + \left(\kappa\right)^2 \, k^2 \,  \left|R'(\kappa)\right| \left[\left<\alpha \right> \left<\chi \right> - \left<\alpha \chi \right> \right].
\end{eqnarray*}
For perturbations to grow over time, and thus drive the steady state~\eqref{eq:barnbarsRcst} unstable, we require at least one root of the  cubic equation~\eqref{eq:cubic} to have a positive real part. Given that $b(k^2)>0$ for all $k^2 \in \mathbb{R}^+$, from the Routh-Hurwitz conditions this can occur if there exist some $k^2 \in \mathbb{R}^+$ for which at least one of the inequalities $c(k^2)<0$, $d(k^2)<0$, or $b(k^2)c(k^2)-d(k^2) <0$ holds. Feasibly, any one of these three inequalities could arise in a suitable parameter region and the algebraic complexity renders statement of a compact single minimal threshold (e.g. as~\eqref{eq:condinstR0beta0}) intractable. We confine here to some general statements. 

Firstly, we note that instabilities in this case could be of either stationary Turing type (in which the eigenvalue that acquires a positive real part is real) or Turing-wave type (in which a complex conjugate pair of eigenvalues acquires a positive real part). 

Secondly, as previously, we observe dependencies on $\left<\alpha \chi \right>$, along with $\left<\alpha \right>$ and $\left<\chi \right>$. This leads us to again expect potentially distinct behaviours according to whether the rate of attractant secretion and the chemotactic sensitivity are positively or negatively correlated, on which we expand further in Section \ref{sec:csgrowth}. Additional insight can be provided by considering some particulars for rates of attractant secretion and chemotactic sensitivities which are monotonic and either positively or negatively correlated. 

For monotonic and positively correlated traits, the Chebyshev integral inequality~\cite[p.~40]{mitrinovic1970analytic} yields $\left< \alpha \chi \right> > \left< \alpha \right>\left< \chi \right>$ and for $d(k^2) < 0$ (recalling the expression~\eqref{eq:wavenumbers} of $k$) we determine the following sufficient condition
\begin{equation}
\label{eq:condinstRcstbeta0a}
\left< \alpha \chi \right> \kappa  > \eta D_n + \frac{m^2 \pi^2}{L^2} D_n D_s + |R'(\kappa)| \, \left(D_s + \eta \frac{L^2}{m^2 \pi^2} \right) \, \kappa \quad \text{for some } \; m \in \mathbb{N}.
\end{equation}
Under monotonic and negatively correlated traits, the Chebyshev integral inequality gives instead $\left< \alpha \chi \right> < \left< \alpha \right>\left< \chi \right>$ and the sufficient condition for instability arising through $d(k^2) < 0$ is
\begin{eqnarray}
\label{eq:condinstRcstbeta0b}
\left< \alpha \chi \right> \kappa  & > & \eta D_n + \frac{m^2 \pi^2}{L^2} D_n D_s + |R'(\kappa)| \, \left(D_s + \eta \frac{L^2}{m^2 \pi^2} \right) \, \kappa + \dfrac{\left(\kappa\right)^2}{D_n} \, \frac{L^2}{m^2 \pi^2} \,  \left|R'(\kappa)\right| \left| \left< \alpha\right>\left< \chi\right>-\left< \alpha\chi\right> \right|  \nonumber \\
&& \text{for some } \; m \in \mathbb{N}.
\end{eqnarray}
Clearly, the condition~\eqref{eq:condinstRcstbeta0b} is more stringent than the condition~\eqref{eq:condinstRcstbeta0a}, reinforcing the notion that the form of trait correlation strongly impacts on pattern formation conditions when phenotype switching is negligible.

While~\eqref{eq:condinstRcstbeta0a} and~\eqref{eq:condinstRcstbeta0b} are sufficient conditions, they do not necessarily provide the minimal critical value of the trait-averaged autoattractant potential, $\left<\alpha \chi \right>$, for pattern formation. Alternatively, denoting the discriminant of the cubic equation~\eqref{eq:cubic} by $\mathcal{D}(k^2)$, that is,
$$
\mathcal{D}(k^2) = b(k)^2 c(k)^2 + 18 b(k) c(k) d(k) - 4 c(k)^3 - 4 b(k)^3 d(k) - 27 d(k)^2,
$$
pattern formation can also occur if there exist some $k^2 \in \mathbb{R}^+$ such that the conditions $c(k^2)<0$, $d(k^2) > 0$, and $\mathcal{D}(k^2) < 0$ simultaneously hold. In fact, this ensures that there exist two complex conjugate roots with positive real part of the cubic equation~\eqref{eq:cubic}. These conditions are clearly cumbersome to study in general but, in appropriate asymptotic regimes, under suitable assumptions on the model functions, it can be shown that, when the rate of attractant secretion and the chemotactic sensitivity are negatively correlated, such conditions provide the minimal critical value of the trait-averaged autoattractant potential for pattern formation.

For instance, if $R'(\kappa) \sim -C$ with $C \in \mathbb{R}^+$ as $\kappa \to \infty$, e.g. when $R(\rho)$ is defined via~\eqref{def:R} and $\gamma \propto \kappa$, then the coefficients of the cubic equation~\eqref{eq:cubic} are such that
\begin{eqnarray*}
&&b \equiv b(k^2) \sim  \kappa \, C, \quad c \equiv c(k^2) \sim \kappa \, k^2 \, \left[\left(D_n + \dfrac{\eta}{k^2} + D_s \right) \, C - \, \left< \alpha \chi \right>\right],
\\\\
&&\text{and } \;\; d \equiv d(k^2) \sim \left(\kappa\right)^2 \, k^2 \,  C  \left(\left< \alpha \right> \left< \chi \right> - \left< \alpha \chi \right> \right) \quad \text{as } \kappa \to \infty.
\end{eqnarray*}
Moreover, the discriminant of the cubic equation~\eqref{eq:cubic} is such that 
\begin{eqnarray*}
\mathcal{D}(k^2) &=& b(k)^2 c(k)^2 + 18 b(k) c(k) d(k) - 4 c(k)^3 - 4 b(k)^3 d(k) - 27 d(k)^2 
\\
&\sim& - 4 \left(\kappa\right)^5 \, k^2 \,  C^4 \,  \left(\left< \alpha \right> \left< \chi \right> - \left< \alpha \chi \right> \right) \quad  \text{as } \kappa \to \infty.
\end{eqnarray*}
Note that $b(k^2)>0$ for all $k^2 \in \mathbb{R}^+$. Again, when the functions $\chi(y)$ and $\alpha(y)$ are monotone in the opposite sense on $\mathcal{Y}$, the Chebyshev integral inequality gives $\left< \alpha \chi \right> < \left< \alpha \right> \left< \chi \right>$. Hence, in this case, $d(k^2)>0$ and $\mathcal{D}(k^2)<0$ for all $k^2 \in \mathbb{R}^+$. Furthermore, if
$$
 \left< \alpha \chi \right> >  \left(D_n + \dfrac{\eta}{k^2} + D_s \right) \, C
$$
then also the condition $c(k^2)<0$ is met. Thus, under this scenario, recalling the expression~\eqref{eq:wavenumbers} of $k$, we find the following sufficient condition for the formation of spatial patterns to occur
\beq
\label{eq:condinstRcstbeta0brhoMinfty}
\left< \alpha \chi \right> > \left(D_s + \eta \frac{L^2}{m^2 \pi^2} + D_n \right) \,  C \quad \text{for some } \; m \in \mathbb{N}.
\eeq
Note that, under this scenario, since $d(k^2) > 0$ for all $k^2 \in \mathbb{R}^+$, the minimal critical value of the trait-averaged autoattractive potential, $\left< \alpha \chi \right>$, for pattern formation to occur is provided by condition~\eqref{eq:condinstRcstbeta0brhoMinfty}.

\subsection{Fast phenotype switching}\label{sec:422}
In the asymptotic regime $\beta \to \infty$, from~\eqref{newsyspertRcst} and~\eqref{newsyspertintconRcst} and following a similar procedure to that employed in Appendix~\ref{sec:annogrbetainfty}, we formally find condition~\eqref{eq:condinstRcstbetainfty} for the formation of spatial patterns to occur, i.e. the condition 
$$
\left<\alpha\right> \left<\chi\right> \kappa  > \eta D_n + \frac{m^2 \pi^2}{L^2} D_n D_s + |R'(\kappa)| \, \left(D_s + \eta \frac{L^2}{m^2 \pi^2} \right) \, \kappa \quad \text{for some } \; m \in \mathbb{N}.
$$

\subsection{Generic phenotype switching}\label{sec:423}
As before, scenarios where $0 < \beta < \infty$ are challenging. However, as in Appendix~\ref{sec:annogrchiconst}, it is again possible to consider the case where $\chi(y)$ is defined via~\eqref{eq:constantchi} and (without loss of generality) set $\mathcal{Y}=(0,1)$. Here, condition~\eqref{newsyspertintconRcst} reduces to
\beq
\label{newsyspertintconRcstred}
\tilde \rho = \tilde{s} \, \dfrac{n^\star \, k^2}{(\lambda + D_n \, k^2 + \left|R'(\kappa)\right| \, \kappa)} \, \chi_0.
\eeq 
Following a similar procedure to that employed in Appendix~\ref{sec:annogrchiconst}, we find condition~\eqref{eq:condinstRcstchiconst} for the formation of spatial patterns to occur, i.e. the condition 
$$
\left< \alpha \right> \, \chi_0 \, \kappa > \eta D_n + \frac{m^2 \pi^2}{L^2} D_n D_s + |R'(\kappa)| \, \left(D_s + \eta \frac{L^2}{m^2 \pi^2} \right) \, \kappa  \quad \text{for some } \; m \in \mathbb{N}.
$$

\bibliographystyle{siam}
\bibliography{refs}

\end{document}